\documentclass[%
 reprint,
 amsmath,amssymb,
 aps,
 prx,
]{revtex4-1}

\usepackage{graphicx}
\usepackage{dcolumn}
\usepackage{bm}
\usepackage{threeparttable}
\usepackage{booktabs}
\usepackage{setspace}
\usepackage{graphicx}
\usepackage{mathtools,amsxtra}
\usepackage{extarrows}
\usepackage[thinlines,thicklines]{easybmat}

\DeclareMathOperator{\sech}{sech}

\begin{document}


\title{Superconducting qubit circuit emulation of a vector spin-1/2}

\author{Andrew J. Kerman}
\affiliation{Lincoln Laboratory, Massachusetts Institute of
Technology, Lexington, MA, 02421}

\date{\today}

\begin{abstract}
We propose a superconducting qubit circuit that can fully emulate a quantum vector spin-1/2, with an effective dipole moment having three independent components whose operators obey the commutation relations of a vector angular momentum in the computational subspace. Each component couples to an independently-controllable external bias, emulating the Zeeman effect due to a fictitious, vector magnetic field, and all three of these vector components remain relatively constant over a broad range of emulated total fields around zero. This capability, combined with established techniques for qubit coupling, should enable for the first time the direct hardware emulation of nearly arbitrary quantum spin-1/2 systems, including the canonical Heisenberg model. Furthermore, it would constitute a crucial step both towards realizing the full potential of quantum annealing, as well as exploring important quantum information processing capabilities that have so far been inaccessible to available hardware, such as quantum error suppression, Hamiltonian and holonomic quantum computing, and adiabatic quantum chemistry.
\end{abstract}

\maketitle

Quantum spin-1/2 models serve as basic paradigms for a wide variety of physical systems in quantum statistical mechanics and many-body physics, and are among the most highly studied in the context of quantum phase transitions and topological order \cite{spinliquids,*kosterlitz,*haldane,*spinice}. In addition, since the spin-1/2 in a magnetic field is one of the simplest realizations of a qubit, many quantum information processing paradigms draw heavily on concepts which originated from or are closely related to quantum magnetism. For example, quantum spin-1/2 language is used to describe nearly all of the constructions underlying quantum error-correction \cite{gottesmanFT,*terhal,*crosspauli} and error-suppression \cite{jiangrieffel,marvianlidar} methods. It is also the most commonly-used framework for specifying engineered Hamiltonians in many other quantum protocols such as quantum annealing \cite{QARMP,*QAfront}, adiabatic topological quantum computing \cite{adiatopo}, quantum simulation \cite{simulation,*bosonsampling,*spinbosonsampling,Dwave3DTFIM,*DwaveKT}, Hamiltonian and holonomic quantum computing \cite{Lloydterhal,*cellularautomata,*hamiltonianQC1D,*holonomic,marvianlidar}, and quantum chemistry \cite{adiaQChem,*bravyikitaev,*molecularenergies}.

The conventional method for simulating vector spin-1/2 Hamiltonians is based on the ``gate-model" quantum simulation paradigm, and uses pulsed, high-speed sequences of discrete, non-commuting gate operations \cite{solanospin,*wallraffspin,*monroespinions} to approximate time-evolution under a desired Hamiltonian \cite{babbushtrotter,*poulintrotter}. In this paradigm, simulating a different Hamiltonian simply requires reprogramming the hardware with a different sequence of gates, a desirable property so long as the error introduced by the discretization can be kept sufficiently low. Unfortunately, this becomes increasingly difficult as the required spin interactions become stronger and/or more complex, since for a fixed gate duration \footnote{There will always be hard practical limit to the frequencies of pulsed gates that can be applied, due to bandwidth limitations of the signal lines, and/or the need to avoid spuriously exciting additional degrees of freedom in the quantum hardware itself.}, the discretization error grows both with the strength of the spin-spin interactions, and with the number of mutually-noncommuting terms they contain. In addition, gate-based implementation necessarily implies that the system occupies Hilbert space far above its ground state, and the resulting information exchange with the environment via both absorption \textit{and} emission is at the root of the need for active quantum error correction \cite{gottesmanFT,*terhal,*crosspauli}.

In light of these considerations, \textit{static} emulation of a desired Hamiltonian (which suffers from neither of the above problems) becomes a potentially appealing alternative for applications requiring strong, complex spin-spin interactions. In fact, the resulting intrinsic ``protection" from noise associated with remaining in the ground state of a quasi-static Hamiltonian is precisely the motivation for adiabatic quantum computation protocols \cite{AQCrev}. To this end, a wide array of what might be called ``weakly-engineered," but still fundamentally naturally-occurring, quantum magnetic systems have been explored for their potential quantum information applications, including, for example: the doped ionic crystals used in the original demonstration of quantum annealing \cite{QA}, ultracold atoms \cite{atomspin} or dipolar molecules \cite{dipolarM}, chemically-engineered molecular nanomagnets \cite{nanomag,*molspin}, and donor spins in Si \cite{Sidonor}. These kind of systems, however, do not have the level of microscopic controllability required in most cases for the applications discussed above.

A game-changing step toward this level of controllability was taken with the advent of the superconducting machines from D-Wave systems \cite{Dwaveflux,Dwavecoupler,Dwave3DTFIM,*DwaveKT}, which are the first examples of large-scale, ``fully-engineered" quantum spin model emulators, and have already been used to great effect in both quantum annealing \cite{QARMP,*QAfront} and quantum simulation \cite{Dwave3DTFIM,*DwaveKT}. However, even these systems have critical limitations in their ability to emulate spin interactions, which are inherited directly from the persistent-current flux qubits on which their spins are based. In particular, as we discuss in detail below, although these qubits are well-suited to emulation of the simpler, transverse-field Ising spin (which interacts with other spins only via its $z$-component), a true vector spin-1/2 is beyond their capabilities. A direct consequence of this is their inability to realize so-called ``non-stoquastic" two-qubit interaction Hamiltonians \cite{bravyistoq}, which have been the subject of intense research interest in recent years \cite{hormoziNS,*henconstrained,*nishimoriNS,*lidarNS,*crossonNS,*albashNS,*araiNS,*terhalNS,gabrielXX} due to their potentially critical role in maximizing the computational power of quantum annealing. In spite of this great potential, no experimental demonstration of such a capability has yet been made, simply because of the limitations of existing qubit hardware.

In this work, we propose a novel superconducting circuit called the ``Josephson phase-slip qubit" (JPSQ), a device which aims for the first time to directly emulate a fully-controllable, quantum vector spin-1/2. When combined with existing techniques for coupling superconducting qubits, this circuit could enable the realization of nearly arbitrary, controllable many-body spin Hamiltonians, without the limitations associated with digital, gate-based quantum simulation methods. Such a capability would provide entirely new modes of access to all of the applications listed above, many of which have never been tried experimentally due to the lack of required capabilities in qubit hardware.

The remainder of the paper is divided as follows: In section ~\ref{s:spinmodel} we begin by describing how persistent-current flux qubits can be used to emulate quantum spin models, and their fundamental limitations in this context. We then introduce in section ~\ref{s:JPSQ} our proposed Josephson phase-slip qubit, analyzing a simplified version of the circuit in detail, and comparing the results with numerical simulations. This section concludes with a discussion of the coherence of the JPSQ in the context of existing superconducting qubit devices. Section ~\ref{s:circuits} contains discussion and simulations of two multi-JPSQ circuit examples: (i) two JPSQs coupled by both $zz$ and $xx$ interactions, and (ii) a four-JPSQ circuit which implements a distance-2 Bacon-Shor logical qubit with quantum error suppression \cite{jiangrieffel,marvianlidar}. Finally, section ~\ref{s:twoisl} describes a generalization of the JPSQ circuit capable of emulating fully-controllable, anisotropic Heisenberg interactions, and we conclude in section~\ref{s:conclusion}.



\begin{figure*}
    \begin{center}
    \includegraphics[width=0.95\linewidth]{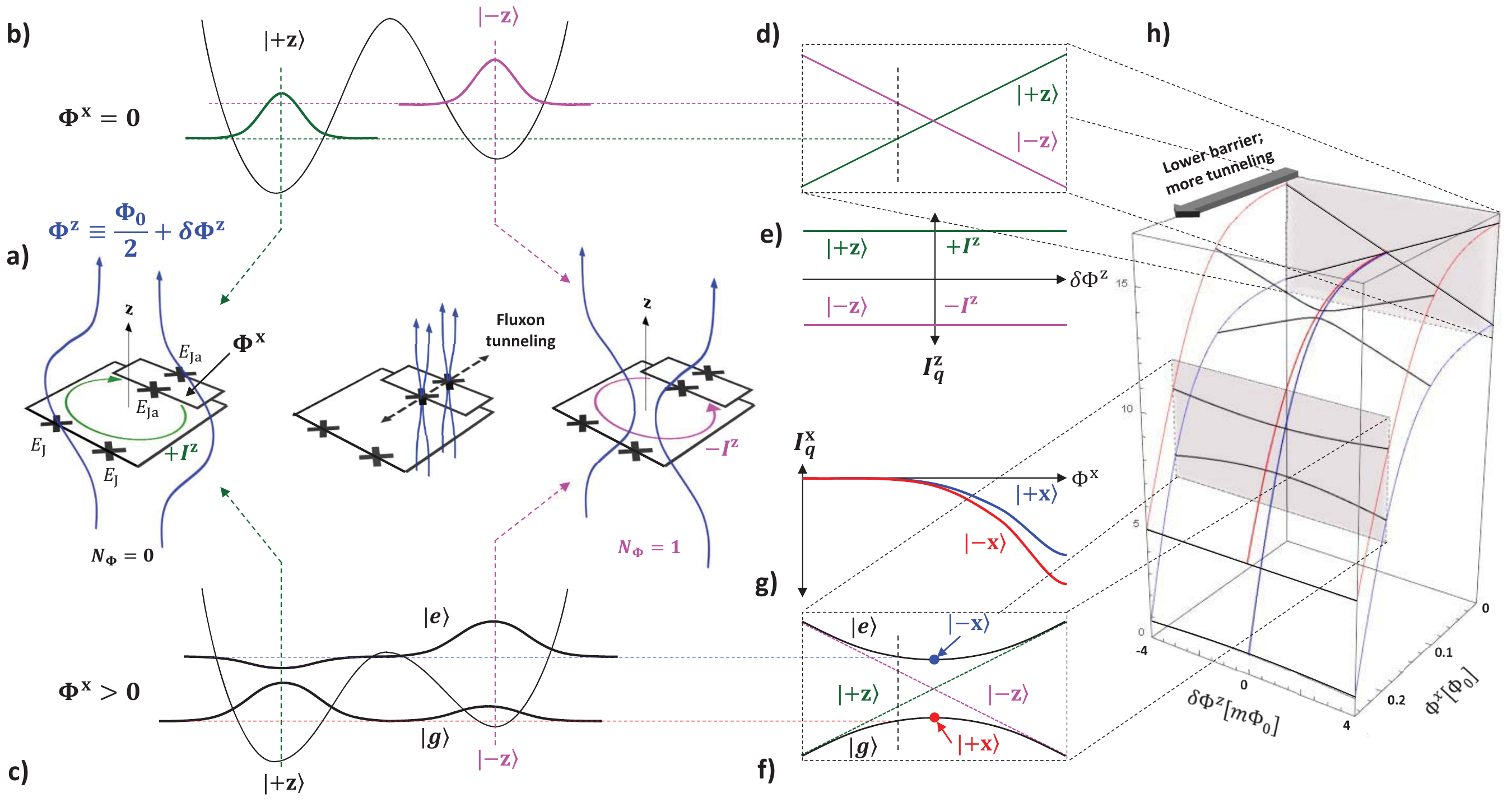}
    \caption[]
        {Spin-1/2 emulation using a two-loop flux qubit. Panel (a) illustrates the persistent current (fluxon) states of the two-loop flux qubit \cite{orlando,mooij,Dwaveflux,MITtunable,fluxmon}. On the left is the state of $N_\Phi=0$ where the persistent current expels the externally-applied flux $\Phi^z$, while on the right is the state $N_\Phi=1$ in which a persistent current in the opposite direction pulls in the additional flux needed to trap exactly one fluxon in the loop. The path connecting these two semiclassical persistent current states to each other traverses an energy barrier in which the fluxon is stored inside the junctions of the DC SQUID interrupting the loop, as illustrated in the center panel of (a). Panels (b) and (c) show how this can be viewed in terms of the motion of a fictitious ``phase" particle in a double-well potential. In (b), the barrier is high, and the two persistent current states are well-isolated, resulting in the energy levels vs. flux shown in (d), which emulate an Ising moment in a $z$ field. Panel (c) illustrates that when the barrier is lowered by threading a flux $\Phi^x$ through the DC SQUID loop, quantum tunneling between the two fluxon states occurs, producing the energy levels in (f). The tunneling appears as an avoided crossing between the two fluxon states of (d), which emulates the effect of a transverse field along $x$. Panels (e) and (g) show the resulting effective persistent currents $I_q^z(\delta\Phi^z)$ and $I^x(\Phi^x)$ for the cases where the emulated transverse field is small, and large, respectively. Finally, panel (h) illustrates the resulting fundamental asymmetry between emulated $z$ and $x$ fields for the 2-loop flux qubit system, for the parameters: $E_{Ja}=h\times44.7$ GHz, $C_{Ja}=1.80$ fF, $E_{J}=h\times134$ GHz, $C_{J}=5.40$ fF (note that the junction capacitances are not shown).}
        \label{fig:fluxqubitfig}
    \end{center}
\end{figure*}

\section{Persistent-current qubits for quantum spin-$\tfrac{1}{2}$ model emulation}\label{s:spinmodel}

Superconducting circuits are already among the most engineerable high-coherence quantum systems available, allowing a range of behavior and interactions to be constructed by design \cite{SCQreview,noriSCreview}. Their capability to emulate complex static spin Hamiltonians is exemplified by the flux-qubit-based machines of D-Wave Systems, Inc. \cite{Dwaveflux,Dwavecoupler,Dwave3DTFIM,*DwaveKT}. These systems are designed to emulate the quantum transverse-field Ising model, with Hamiltonian:

\begin{equation}
\hat{H}_\textrm{TIM}=-\sum_i(E^z_i\hat{\sigma}^z_i+E^x_i\hat{\sigma}^x_i)-\sum_{i<j}J_{ij}\hat{\sigma}^z_i\hat{\sigma}^z_j\label{eq:TIM}
\end{equation}

\noindent Each spin has an emulated local magnetic field in the $x-z$ plane with effective Zeeman energies given by the parameters $E^z_i$ and $E^x_i$ in eq.~\ref{eq:TIM}, and pairwise couplings to other spins parametrized by the Ising interaction energies $J_{ij}$. Notably absent from this Hamiltonian are any interactions between the transverse moments of the spins, whose physical realization is the subject of this work. To better understand what follows, we first describe the physics underlying emulation of eq.~\ref{eq:TIM} with 2-loop flux qubits \cite{orlando,mooij,Dwaveflux,MITtunable,fluxmon}, and why these circuits cannot be used to emulate the vector spin interactions of interest here.

Figure~\ref{fig:fluxqubitfig} illustrates how flux qubits are currently used to emulate quantum spins. The circuit shown in panel (a) is the basic 4-junction flux qubit \cite{orlando,mooij,Dwaveflux,MITtunable,fluxmon}, having two loops biased by fluxes $\Phi^z$ and $\Phi^x$, labelled according to the spin moment they are used to emulate \footnote{Note that we focus here on the flux qubit variant which uses Josephson junctions for its loop inductance \cite{orlando}, instead of the RF-SQUIDs used in quantum annealing applications to date \cite{Dwaveflux,Dwavecoupler,fluxmon}; however, the physics is qualitatively the same, and the conclusions presented here apply equally to either type.}. When the $z$ loop is biased with an external flux $\Phi^z=\Phi_0/2+\delta\Phi^z$ (where $\Phi_0\equiv h/2e$ is the superconducting fluxoid quantum, and $\delta\Phi^z\ll\Phi_0$), the two lowest-energy semiclassical states of the loop are nearly degenerate due to the Meissner effect, having approximately equal and opposite supercurrents. As shown in fig.~\ref{fig:fluxqubitfig}(a), these two semiclassical states correspond to expulsion of the external flux from the loop, or pulling additional flux into it, such that it contains exactly zero or one fluxoid quantum, respectively. They can be identified with two local minima in a magnetic potential (panel (b) in the figure) experienced by a fictitious particle whose ``position" corresponds to the gauge-invariant phase difference across the two larger Josephson junctions (which play the role of a loop inductance), and whose ``momentum" corresponds to the total charge that has flowed through them. The difference in potential energy between these two minima is controlled by $\delta\Phi^z$ (panel (d)), and approximately corresponds to the interaction energy between an applied field and the two equal and opposite persistent currents (panel (e)). These two states naturally play the role of $|\pm z\rangle$, the eigenstates of $\hat{\sigma}^z$ for the emulated spin. Quantum coupling between these two states corresponds to the operator $\hat{\sigma}^x$: the Zeeman Hamiltonian due to an emulated transverse field. As shown in (a)-(c), this can be viewed as tunneling of a ``fluxon" between the two states in which it is inside or outside of the larger loop, with a strength controlled by the barrier height via $\Phi^x$ (panel (c)).

In published experimental work to date (with the exception of ref.~\onlinecite{gabrielXX}), this $\Phi^x$ tuning has only been used to adjust the Zeeman energy due to an effective transverse field \cite{orlando,mooij,Dwaveflux,MITtunable,fluxmon}. However, we are concerned here with producing a static interaction between the transverse dipole moments of two such emulated spins. Figure~\ref{fig:xxflux}(a) shows the circuit which realizes such an interaction between two-loop flux qubits, originally proposed in ref.~\onlinecite{AJKsim}, and demonstrated in ref.~\onlinecite{gabrielXX}. The DC SQUID loops of the two qubits (whose external flux biases control the height of their fluxon tunnel barriers) are each coupled by a mutual inductance $M$ to a common flux qubit coupler. The latter can be viewed semiclassically as a tunable effective inductance $L_C^\textrm{eff}$ which can assume positive or negative values, depending on how it is biased \cite{plourde,niskanen,niskanen2,ashhab,Dwavecoupler,MITLLcoupled}; therefore, we can qualitatively understand the resulting two-qubit interaction using the Hamiltonian:

\begin{figure*}
    \begin{center}
    \includegraphics[width=0.9\linewidth]{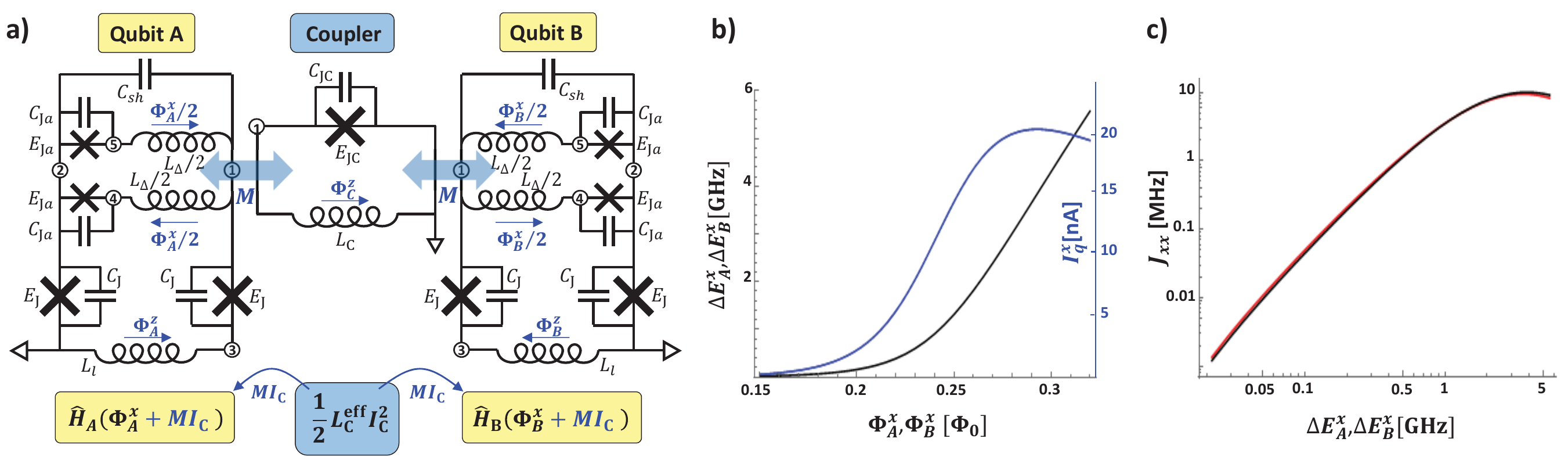}
    \caption[]
        {Transverse interaction between two-loop flux qubits, used to emulate an $xx$ interaction. Panel (a) shows a circuit which realizes such an interaction, as in refs.~\onlinecite{AJKsim,gabrielXX}. The two qubits' DC SQUID loops are coupled magnetically to a common flux qubit coupler, having effective inductance $L_C^\textrm{eff}$. Panel (b) shows the a full simulation \cite{JJcircuitSim} of the effective $x$-field Zeeman energy $\Delta E^x$ (left axis) and $I_q^x$ (right axis) vs. $\Phi^x$ for a single qubit. The parameters used are: $E_{J}=h\times117$ GHz, $C_J=4.42$ fF, $E_{Ja}=h\times47.8$ GHz, $C_{Ja}=1.80$ fF, $L_l=110$ pH, $L_\Delta=20$ pH, $C_\textrm{sh}=35$ fF. Panel (c) shows the resulting two-qubit coupling energy $J_{xx}$ as a function of $\Delta E^x$, where the black line is obtained from eq.~\ref{eq:Jxx} using the $I_q^x$ from (b), $M=40$ pH, and $L_C^\textrm{eff}=-25$ pH from numerical simulation with: $E_{JC}=h\times376$ GHz, $C_{JC}=14.2$ fF, and $L_C=575$ pH. For comparison, the red line shows the $J_{xx}$ obtained from numerical simulation of the full three-qubit system \cite{JJcircuitSim}. Because of the exponential decrease of $I_q^x$ with $\Phi^x$ shown in (b), the coupling energy goes exponentially to zero with $\Delta E^x$, so that non-negligible transverse interaction can only be achieved in the presence of large offset $x$ fields for both qubits.}
        \label{fig:xxflux}
    \end{center}
\end{figure*}

\begin{eqnarray}
\hat{H}&=&\hat{H}_A(\Phi_A^x+MI_C)+\hat{H}_B(\Phi_B^x+MI_C)+\frac{1}{2}L_C^\textrm{eff}I_C^2\nonumber\\
&\equiv&-E^x_A\hat\sigma^x_A-E^x_B\hat\sigma^x_A-J_{xx}\hat\sigma^x_A\hat\sigma^x_B\label{eq:Hxx}
\end{eqnarray}

\noindent where the coupler ground state is described semiclassically in terms of its inductance $L_C^\textrm{eff}$ and current $I_C$, $\Phi_A^x,\Phi_B^x$ are the static flux offsets applied to the two qubits' DC SQUID loops, and we describe each qubit $q\in\{A,B\}$ in terms of its $\Phi^x-$dependent quantum eigenenergies (in the absence of coupling) at zero effective $z$ field ($\Phi^z=\Phi_0/2$):

\begin{equation}
\hat{H}_q(\Phi^x_q)\equiv E_q^{+x}(\Phi^x_q)|+x\rangle\langle+x|+E_q^{-x}(\Phi^x_q)|-x\rangle\langle-x|
\end{equation}

\noindent To calculate the effective coupling energy $J_{xx}$ in eq.~\ref{eq:Hxx}, we expand the qubit energies around the points $\Phi_A^x$ and $\Phi_B^x$, and then minimize the total energy with respect to the coupler current $I_C$ (following ref.~\onlinecite{AJKlongit}), to obtain:

\begin{equation}
J_{xx}=I_A^xI_B^x\frac{M^2}{L_C^\textrm{eff}}\frac{1}{1+\tfrac{M^2}{L_C^\textrm{eff}}\tfrac{L_A+L_B}{L_AL_B}}\label{eq:Jxx}
\end{equation}

\noindent where we have defined the Taylor coefficients (with $q\in\{A,B\}$):

\begin{eqnarray}
I_q^x&\equiv&\frac{d}{d\Phi_q^x}\Delta E_q^x,\;\;\;\Delta E_q^x\equiv\frac{E_q^{-x}-E_q^{+x}}{2}\label{eq:Ip}\\
L_q^{-1}&\equiv&\frac{d^2}{d(\Phi_q^x)^2}\bar{E}_q^x,\;\;\;\bar{E}_q^x\equiv\frac{E_q^{-x}+E_q^{+x}}{2},
\end{eqnarray}

\noindent and neglected the differential quantum inductance between the $|\pm x\rangle$ states. Equations \ref{eq:Jxx} and \ref{eq:Ip} show that the semiclassical quantity which plays the role of the qubit magnetic moment along the fictitious $x$ direction is $I_q^x$, the slope of the tunnel splitting energy $\Delta E_q^x$ with respect to $\Phi^x$. This quantity is plotted in fig.~\ref{fig:xxflux}(b) in blue (right axis) as a function of $\Phi^x$ for a single tunable flux qubit, using the full numerical simulation methods of ref.~\onlinecite{JJcircuitSim} (previously used in refs.~\onlinecite{MITLLflux,MITLLcoupled}). Panel (c) shows in black the corresponding transverse coupling energy $J_{xx}$ obtained by plugging this into eq.~\ref{eq:Jxx}. The crucial point here is that the $xx$ coupling that can be achieved in this way is always much smaller than the local transverse-field Zeeman energies: $J_{xx}\ll \Delta E_A^x,\Delta E_B^x$. The physical reason for this is simple: the $\Phi^x_q$-dependence of the tunneling energy $\Delta E^x_q$ is exponential, since increasing the flux corresponds to lowering the tunnel barrier [c.f., fig.~\ref{fig:fluxqubitfig}(h)]. Since the effective $x$ magnetic moment of each qubit [c.f.,~eq.~\ref{eq:Ip}] is the derivative with respect to $\Phi^x$ of this energy, it can only be large when the energy is itself large. In spin language, this constraint on transverse coupling between flux qubits corresponds to the $x$ magnetic moments of the emulated spins going exponentially to zero as their local $x$ fields go to zero, as shown in fig.~\ref{fig:fluxqubitfig}(g). This limitation renders present-day superconducting circuits unable to emulate (in the static fashion under discussion here) the majority of the spin models discussed in the introduction, with the notable exception of the transverse field Ising model of eq.~\ref{eq:TIM} emulated by D-Wave systems' machines \cite{Dwave3DTFIM,*DwaveKT}.

\begin{figure*}
    \begin{center}
    \includegraphics[width=0.9\linewidth]{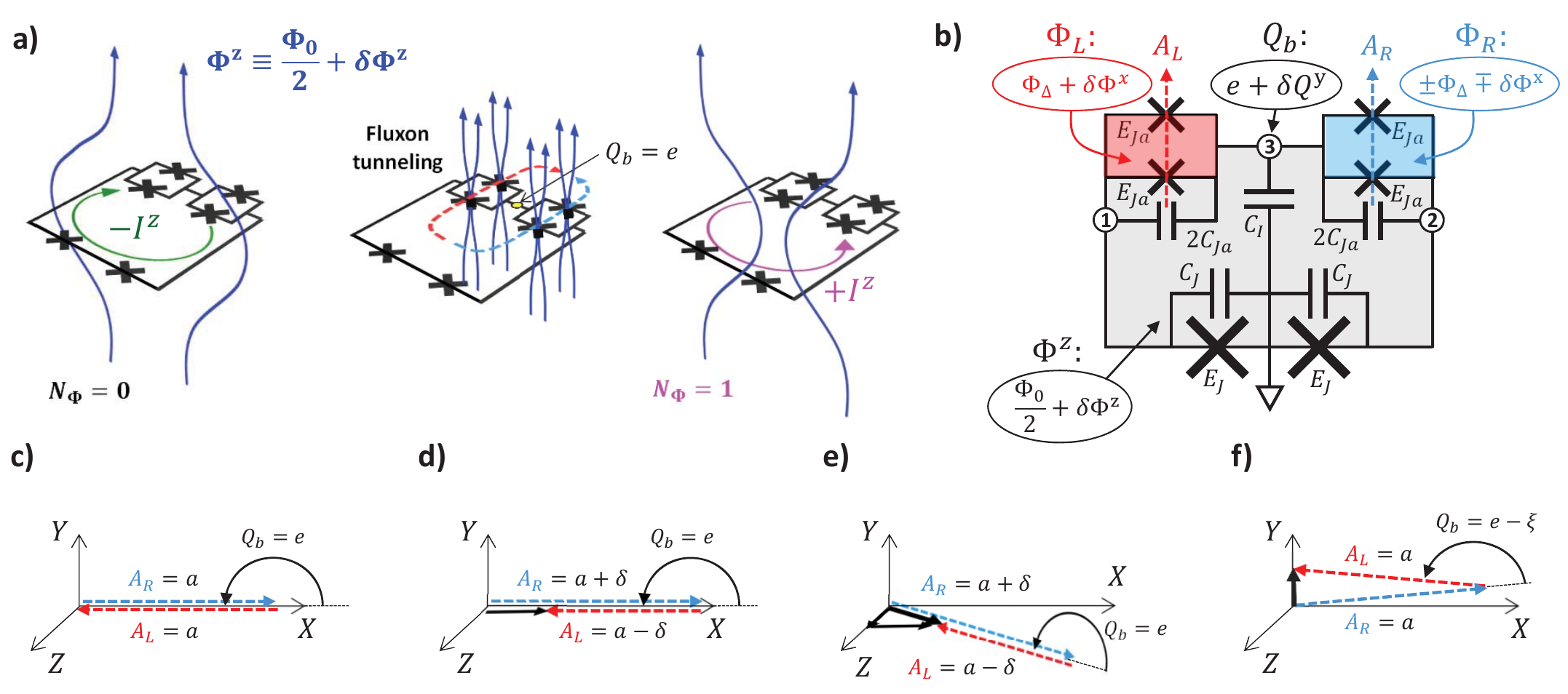}
    \caption[]
        {Circuit for emulation a vector spin: the Josephson phase-slip qubit. Panel (a) illustrates the persistent current (fluxon) states of the proposed circuit, and the two fluxon tunneling paths into or out of the loop. On the left is the state of $N_\Phi=0$ where the persistent current expels the externally-applied flux $\Phi^z$, while on the right is the state $N_\Phi=1$ in which a persistent current in the opposite direction pulls in the additional flux needed to trap one fluxon in the loop. Two paths connect these two semiclassical persistent current states to each other, each of which contains an energy barrier (in which the fluxon is stored inside the junctions of one of the two DC SQUIDs interrupting the loop) as illustrated in the center panel of (a). Panel (b) shows the simplified circuit considered in our analytic analysis. Panels (c)-(f) illustrate the fluxon tunneling amplitudes which create an effective transverse-field Zeeman splitting, under several different conditions: in (c), the island is polarized with an offset charge $e$, and the two DC SQUIDs are biased with flux of equal magnitude $\Phi_\Delta$, corresponding to zero effective field; in (d), the two DC SQUID fluxes are imbalanced such that the two fluxon tunneling amplitudes no longer cancel completely, corresponding to a nonzero effective $\pm x$ field; in (e) a nonzero $z$ field is added; finally, in (f) the island polarization charge is displaced by $\xi$, resulting in a nonzero effective $y$ field.}
        \label{fig:JPSQfig1}
    \end{center}
\end{figure*}

\section{Josephson phase-slip qubit (JPSQ)}\label{s:JPSQ}

In section~\ref{s:spinmodel}, we described how emulation of a large $x$ magnetic moment with a persistent-current qubit requires its tunneling energy to be sensitive to barrier height, which implies that there must be substantial probability inside the barrier. For this to remain true all the way to zero transverse field (tunneling energy), the wavefunction inside the barrier must remain appreciable \textit{even when the tunneling itself goes to zero}, a self-contradictory requirement that cannot be satisfied by conventional flux qubits. Figure~\ref{fig:JPSQfig1} shows a persistent-current qubit circuit which can. Here, the tunable fluxon tunnel barrier that provides control of the transverse field (which in fig.~\ref{fig:fluxqubitfig}(a) is realized with a DC SQUID) consists of \textit{two} DC SQUIDs in series, separated by a central superconducting island whose polarization charge can be controlled with an external bias voltage. This object is similar to the so-called ``quantum phase-slip transistor" (QPST) \cite{hongisto,*hriscu,*AJKdual,*cQUID} (with the quantum phase-slip junctions here replaced by DC SQUIDs), hence the name ``Josephson phase-slip qubit" (JPSQ). The key feature which motivates the use of a QPST to control fluxon tunneling in the present context is that it provides \textit{two} fluxon tunneling paths into or out of the loop, as illustrated in (a). If a polarization charge $Q_b$ is present on the island, the two fluxon tunneling amplitudes acquire a relative phase shift due to the Aharonov-Casher effect \cite{aharonovcasher}; when $Q_b=e$ this phase shift is $\pi$, and if the magnitudes of the two tunneling amplitudes are equal, total suppression of fluxon tunneling occurs and the transverse field Zeeman energy is zero [fig.~\ref{fig:JPSQfig1}(c)]. Crucially, this remains true \textit{even when the individual tunneling amplitudes are large and flux-sensitive}, allowing a strong, linear flux-sensitivity (magnetic dipole moment) to persist even around zero field. We note before proceeding that a number of previous works have highlighted and/or experimentally exploited this phenomenon as a means to observe the Aharonov-Casher effect in superconducting circuits \cite{ACaverin,*glazman,*manucharyan,*guichard,*bell}; here, we propose a way to use it to realize a superconducting-circuit-based vector spin-1/2 qubit.

Figure ~\ref{fig:JPSQfig1}(b) shows a simplified JPSQ circuit in more detail, and in particular how a transverse magnetic moment can be realized. The two DC SQUIDs are biased with an offset flux of the same magnitude $\Phi_L=\pm\Phi_R=\Phi_\Delta$, so that their individual fluxon tunneling amplitudes $A_L(\Phi_\Delta)$ and $A_R(\pm\Phi_\Delta)$ have equal magnitude, and are flux-sensitive (the $\pm$ indicates that there are two possible choices for the relative sign). If we then magnetically couple an input flux $\delta\Phi^x$ to both DC SQUIDs with equal strength, and signs such that the resulting total flux through the two DC SQUIDs is affected oppositely, the two tunneling amplitudes no longer cancel, creating an effective transverse field as illustrated in fig.~\ref{fig:JPSQfig1}(d). If we also change the flux $\delta\Phi^z$, the total effect is analogous to a field in between the $x$ and $z$ axes, as shown in Panel (e). Panel (f) shows the effect of charge displacements away from half a Cooper pair, which act as transverse fields in the $y$ direction.

\subsection{JPSQ Hamiltonian}\label{sub:Ham}

We now validate the intuitive picture given above by analyzing the quantum mechanics of the JPSQ circuit. As we show below, this analysis can be used both to understand the physics of the circuit, and to make semi-quantitative predictions of its important properties. The classical Hamiltonian for the JPSQ circuit of fig.~\ref{fig:JPSQfig1}(b) can be written:

\begin{equation}
H=\frac{1}{2}\vec{Q}^T\cdot\mathbf{C}^{-1}\cdot\vec{Q}-\vec{Q}_b\cdot\vec{V}+U_J\label{eq:JPSQH}
\end{equation}

\noindent where the three terms are: (i) the electrostatic energy; (ii) the interaction with a polarization charge $\vec{Q}_b=\left(0,0,Q_b\right)$ (supplied by a bias source); and (iii) the Josephson potential energy. We start by transforming to the following loop, island, and plasma mode phase coordinates:

\begin{eqnarray}
\phi_l&\equiv&\phi_2-\phi_1\nonumber\\
\phi_I&\equiv&\phi_3+\tfrac{1}{2}\left(\phi_\Delta-\delta\phi^z-\delta\phi^x\right)\nonumber\\
\phi_p&\equiv&\tfrac{1}{2}\left(\phi_1+\phi_2\right)
\end{eqnarray}

\noindent where here, and below, we use lowercase $\phi$ to indicate dimensionless flux (phase) quantities, according to: $\phi\equiv2\pi\Phi/\Phi_0$. In this coordinate representation, the Josephson potential is, to first order in $\delta\Phi^x$:

\begin{widetext}
\begin{eqnarray}
\frac{U_J(\phi_I,\phi_l,\phi_p)}{2\tilde{E}_{Ja}(\phi_\Delta)}=\beta(\phi_\Delta)\left[1-\cos{\frac{\phi_l}{2}}\cos{\phi_p}\right]-\cos\frac{\delta\phi^x}{2}\sin{\left(\phi_I-\phi_p\right)}\sin{\left(\frac{\delta\phi^z-\phi_l}{2}\right)}\hspace{2cm}\nonumber\\
+\sec\frac{\phi_\Delta}{2}\left[1-\sin\frac{\delta\phi^x}{2}\sin\frac{\phi_\Delta}{2}\cos{\left(\phi_I-\phi_p\right)}\cos{\left(\frac{\delta\phi^z-\phi_l}{2}\right)}\right]
\label{eq:JPSQpot}
\end{eqnarray}
\end{widetext}

\begin{figure*}
    \begin{center}
    \includegraphics[width=1.0\linewidth]{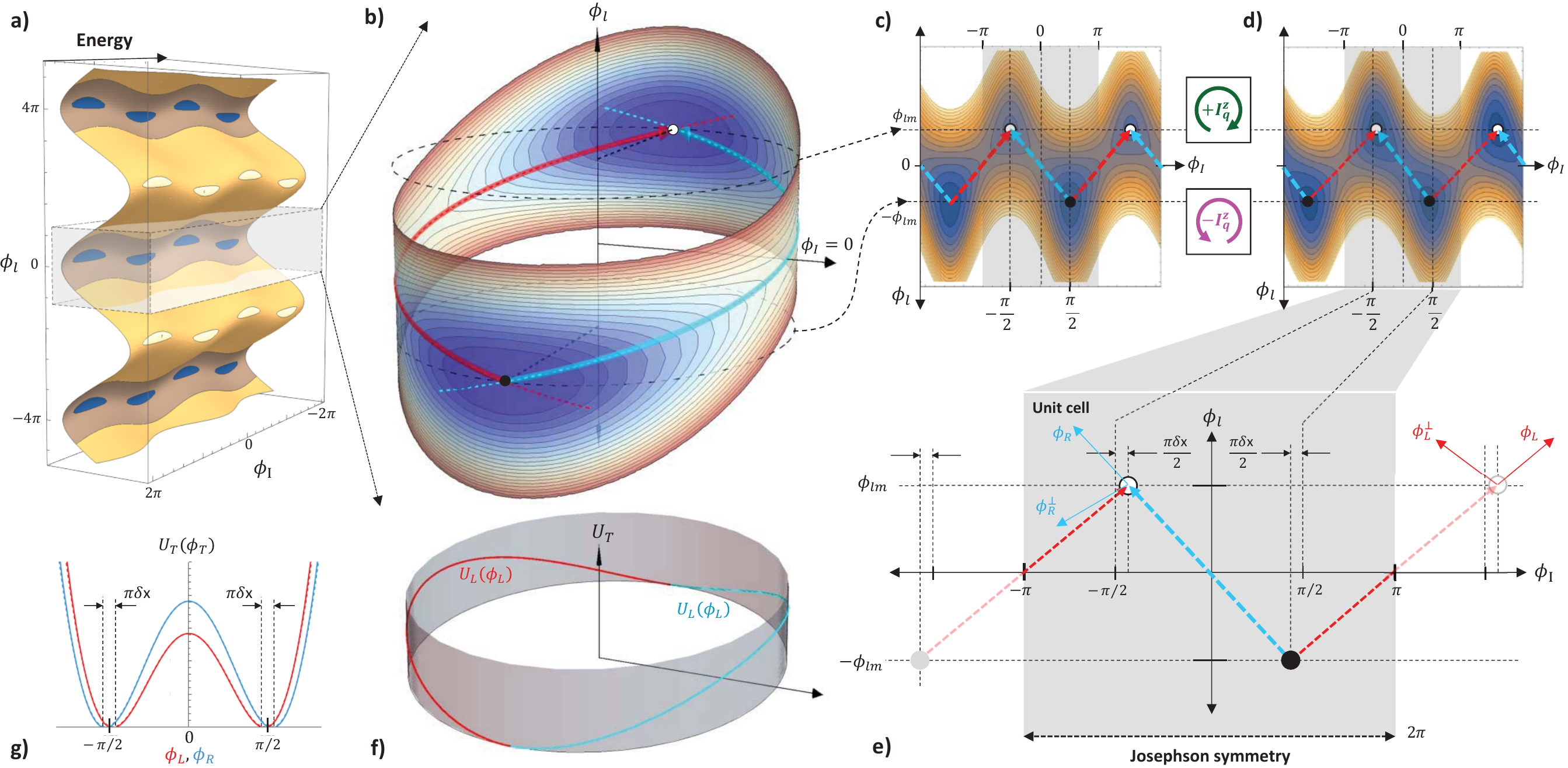}
    \caption[]
        {Josephson potential for the JPSQ. Panel (a) shows a cut through potential of the circuit in fig.~\ref{fig:JPSQfig1}(b) obtained by setting $\phi_p=0$ (a very good approximation for the parameters of interest here). Its two coordinates are: $\phi_l$, the phase difference across the two larger loop junctions; and $\phi_I$, the phase of the island. Since the Josephson barrier along $\phi_l$ (for tunneling of fluxons through the two big junctions) is much larger than other energy scales, we neglect tunneling through it and focus on the region near $\phi_l=0$, illustrated by the shaded box in (a). Panel (b) shows a contour plot of the potential in this reqion, where the Josephson translational symmetry along $\phi_I$ has been made explicit by treating it as an angular coordinate. Inside each unit cell of this potential (one angular period of $\phi_I$) there are two local minima, corresponding to the two persistent current states having equal and opposite values of $\phi_l$ (indicated by dashed circles). Red and blue 3D arrows indicate the two separate paths that connect these two minima (associated with the two possible angular directions in $\phi_I$), which correspond to tunneling of a fluxon through one or the other of the two DC SQUIDs in fig.~\ref{fig:JPSQfig1}(b). Panels (c) and (d) compare the potential for the cases: $\delta\phi^z\neq0,\delta\phi^x=0$ and $\delta\phi^z=0,\delta\phi^x\neq0$, respectively. To facilitate this comparison, they are shown ``unwound" along the $\phi_I$ coordinate, with two full periods visible, and the persistent current states labelled by horizontal dashed lines. From this perspective, the two tunneling paths correspond to inter-unit-cell and intra-unit-cell tunneling (the unit cell is indicated with gray shading). Panel (e) illustrates quantities defined in the text, for a single unit cell: thin, solid red and blue arrows illustrate the coordinates defined in eq.~\ref{eq:coords}; thick, dashed, red and blue arrows the linear approximation to the extremal paths used in our analysis; and, thin black arrows the displacements along $\phi_I$ of the two potential minima when $\delta x\neq0$. Panel (f) illustrates the approximate 1D potential, defined on a circle, obtained by evaluating the full potential on the paths shown in (d). Panel (g) shows the potentials for the two tunneling paths overlaid (solid lines - exact result from eq.~\ref{eq:JPSQpot}, dashed lines - eq.~\ref{eq:k6}), illustrating that when $\delta x\neq0$, the barrier and the length of one path are increased while for the other they are decreased.}
        \label{fig:JPSQpot}
    \end{center}
\end{figure*}

\noindent where we have defined the effective DC SQUID Josephson energy $\tilde{E}_{Ja}(\phi_\Delta)\equiv 2E_{Ja}\cos{(\phi_\Delta/2)}$ and the ratio: $\beta(\phi_\Delta)\equiv E_{J}/\tilde{E}_{Ja}(\phi_\Delta)$ between the Josephson inductances of the DC SQUIDs (the fluxon tunneling elements) and that of the larger loop junctions (functioning as the inductance appearing ``across" them) \footnote{This parameter is analogous to the Stewart-McCumber parameter of an RF SQUID, and is also related to the quantity $\alpha$ which is conventionally used to describe 3- and 4-junction flux qubits by: $\alpha=1/2\beta$}. In this circuit, just as in the case of 3- and 4-junction flux qubits, the plasma mode $\phi_p$ (the symmetric oscillation across the two larger Josephson junctions), can in most cases of interest be treated as a ``bystander," in the sense that the energy barrier between fluxon states along this direction, as well as its characteristic oscillation frequency, are both usually much larger than the energy scale of interest for the qubit, such that to a good approximation the $\phi_p$ mode does not participate in relevant phenomena at that energy scale.

Figures~\ref{fig:JPSQpot}(a)-(b) show the potential energy surface of eq.~\ref{eq:JPSQpot} obtained by setting $\phi_p=0$, for $\delta\phi^z=\delta\phi^x=0$. Whereas the essential physics of a conventional flux qubit can be approximately described in terms a phase particle moving in a one-dimensional double-well potential [c.f., figs.~\ref{fig:fluxqubitfig}(b),(c)] \cite{orlando}, whose position corresponds to the gauge-invariant phase across the qubit's loop inductance, the JPSQ is fundamentally different in that at least two dynamical variables are needed to capture even its qualitative properties. At a very high level, this difference can be associated with the fact that Josephson symmetry (under translations of $\Phi_0$) does not play an essential role in the important low-energy properties of the flux qubit, while it is essential to those of the JPSQ: for the flux qubit of fig.~\ref{fig:fluxqubitfig}(a), if we neglect spurious fluxon tunneling through the larger, loop junctions, the only nontrivial tunneling path for a fluxon is the usual one connecting the two persistent current states. However, as shown in figs.~\ref{fig:JPSQfig1} and ~\ref{fig:JPSQpot}, for the JPSQ there is a \textit{closed} fluxon tunneling path which encircles the island, corresponding to a discrete Josephson symmetry. This symmetry is highlighted in fig.~\ref{fig:JPSQpot}(b) by representing $\phi_I$ (the phase of the island) as an angular coordinate \footnote{In this sense the JPSQ, although it is intended to function as a persistent-current flux qubit, has something in common with the well-known charge \cite{charge,cQED} (or transmon \cite{transmon,*DiCarlotransmon,*martinistransmon}) and quantronium \cite{quantronium} qubits, whose physics at a high level is that of a superconducting island with Josephson coupling to ground. As illustrated in fig.~\ref{fig:JPSQpot}, the JPSQ in fact combines elements of these two classes of superconducting qubits (charge and flux), usually viewed as qualitatively distinct. From a flux qubit point of view, the JPSQ exhibits a double-well potential in the form of two weakly-coupled 1D periodic potentials (assuming tight binding); from a charge qubit point of view, the JPSQ exhibits a 1D periodic Josephson potential whose unit cell contains a double-well. In fact, the JPSQ can be \textit{continuously varied} between regimes of flux qubit and charge qubit behavior by changing its design and bias parameters.}.

Fig.~\ref{fig:JPSQpot}(b) illustrates how the two fluxon tunneling paths between persistent current states shown in fig.~\ref{fig:JPSQfig1}(a) appear on the 2D potential surface, as red and blue arrows, and how they correspond to motion in the two angular directions along $\phi_I$. This coordinate can be viewed as the angular coordinate of a fluxon circling around the island (passing through the DC SQUIDs' junction barriers), one period of which corresponds to a voltage pulse on the island of area $\pm\Phi_0$, with the sign determined by the direction of motion. When there is an offset charge on the island, the system becomes sensitive to this direction, via the second term of eq.~\ref{eq:JPSQH}; when the island is polarized with exactly half a Cooper pair, the resulting $\pi$ relative phase shift produces the destructive interference shown in fig.~\ref{fig:JPSQfig1}(c). We will see below that this can also be viewed as a geometric phase shift.

The two local minima of eq.~\ref{eq:JPSQpot}, illustrated in fig.~\ref{fig:JPSQpot}(b), correspond semi-classically to the two persistent current states, and are given by (for $\phi^z=\pi$, and to first order in $\delta\phi^x$):

\begin{eqnarray}
\left(\phi_I,\phi_l,\phi_p\right)&=&
\begin{cases}
\left(-\phi_{Im},-\phi_{lm},0\right)\\
\left(\phi_{Im},\phi_{lm},0\right)
\end{cases}\nonumber\\
\phi_{Im}&\equiv&\frac{\pi}{2}-\delta\phi^x\frac{\cot\theta}{2}\tan\tfrac{\phi_\Delta}{2}\nonumber\\
\phi_{lm}&\equiv&2\theta\label{eq:minima}
\end{eqnarray}

\noindent where we have defined: $\theta(\phi_\Delta)\equiv\cot^{-1}[\beta(\phi_\Delta)]$. In the limit where quantum phase fluctuations about these classical minima and the tunneling between them are negligible, the average magnitude of the corresponding equal and opposite persistent supercurrents circulating in the loop can be written down directly:

\begin{equation}
I_{\textrm{JPSQ}}^z\approx\tilde{I}_{Ca}(\phi_\Delta)\cos[\theta(\phi_\Delta)]\label{eq:IpJPSQ}
\end{equation}

\noindent where $\tilde{I}_{Ca}(\phi_\Delta)\equiv \tilde{E}_{Ja}(\phi_\Delta)\times2\pi/\Phi_0$ is the effective critical current of each DC SQUID. The corresponding result for the semi-classical persistent current of the flux qubit is:

\begin{figure}
    \begin{center}
    \includegraphics[width=0.9\linewidth]{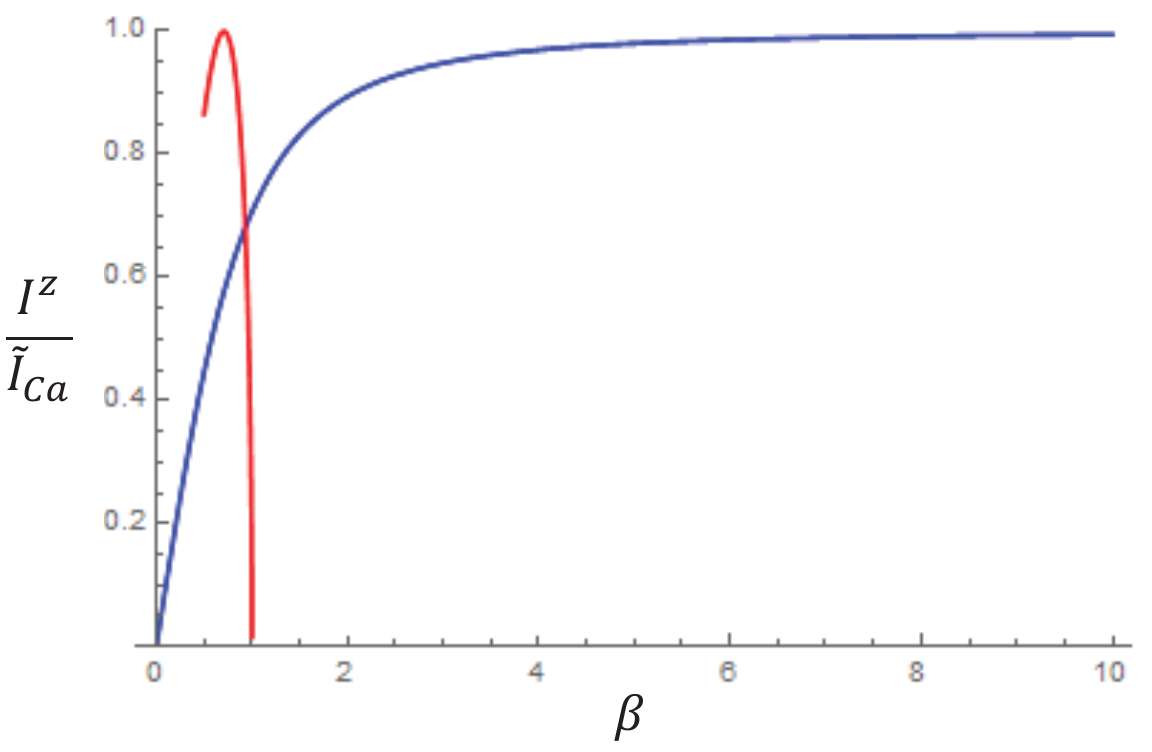}
    \caption[]
        {Semiclassical persistent currents of the JPSQ (blue line) and flux qubit (red line). The horizontal coordinate $\beta$ is the ratio between the Josephson inductances of the fluxon tunneling element(s) and the loop (see text). For the tunable flux qubit, this quantity is related to the well-known parameter $\alpha$ according to: $\alpha=1/2\beta$, such that the excluded region $\beta<0.5$ corresponds to $\alpha>1$.}
        \label{fig:fluxcomp}
    \end{center}
\end{figure}

\begin{equation}
I_{\textrm{flux}}^z\approx\tilde{I}_{Ca}(\phi_\Delta)\sin[2\gamma(\phi_\Delta)]\label{eq:Ipflux}
\end{equation}

\noindent where the angle $\gamma(\phi_\Delta)$ is defined by: $2\cos[\gamma(\phi_\Delta)]\equiv\beta(\phi_\Delta)$. Equations~\ref{eq:IpJPSQ} and~\ref{eq:Ipflux} are plotted in fig.~\ref{fig:fluxcomp}, and they exhibit a qualitative difference between the two circuits when viewed as persistent-current qubits. The flux qubit result is shown in red, and displays the well-known behavior that a double-well potential only exists when $0.5<\beta<1$; that is, the Josephson inductance of the small junction must be less than that of the loop. Within this range, the persistent current varies strongly, having a peak at $\beta=1/\sqrt{2}$. By constrast, the JPSQ can have a well-defined classical persistent current for any value of $\beta$, and in fact the current asymptotically approaches $\tilde{I}_{Ca}$ as $\beta$ gets larger; that is, as the loop inductance becomes negligible compared to that of the DC SQUIDs. Note that this is a regime where the flux qubit does not have a double-well potential at all.

This qualitative difference can be intuitively explained by examining the effective inductive division occurring in the two cases. For the JPSQ, the top of the potential barrier between persistent current states corresponds to a gauge-invariant phase difference of $\pi/2$ across each of the two DC SQUIDs, such that their effective Josephson inductances formally diverge. The SQUIDs therefore look approximately like two series current sources, supplying their critical current $\tilde{I}_{Ca}(\phi_\Delta)$ to a ``load" consisting of the two larger junctions (as long as $\beta$ is not close to 1). For the tunable flux qubit, however, the top of the potential barrier between persistent current states corresponds to a gauge-invariant phase difference of $\pi$ across the (single) DC SQUID, which is its point of \textit{minimum} (and negative) Josephson inductance. The result is that unlike the JPSQ, the persistent current of the flux qubit is determined by a balance between the Josephson inductance of the DC SQUID and that of the two loop junctions in series, which are of similar magnitude. Only near the point $\beta=1/\sqrt{2}$, where the potential minima correspond to $\pm\pi/2$ across the DC SQUID (and where its Josephson inductance diverges) does the persistent current approach $\tilde{I}_{Ca}(\phi_\Delta)$.

\subsection{Analysis of fluxon tunneling in the JPSQ}\label{sub:tunnel}

We now discuss tunneling between the JPSQ's persistent current states, denoted here by $|+z\rangle$ and $|-z\rangle$, using the Euclidean path-integral representation for the transition amplitude between them:

\begin{eqnarray}
\langle -z|e^{-\frac{\hat{H}\tau}{\hbar}}|+z\rangle=\oint\mathcal{D}[\vec\phi(\tau)]e^{-\mathcal{S}_E[\vec\phi(\tau)]}\label{eq:T}
\end{eqnarray}

\noindent Here, $\mathcal{D}[\vec\phi(\tau)]$ indicates a functional integral over all paths $\vec\phi(\tau)=\left(\phi_I(\tau),\phi_l(\tau)\right)$ connecting the classical minima of the two potential wells, and $\mathcal{S}_E[\vec\phi(\tau)]$ is the corresponding Euclidean action associated with each such path. Anticipating the validity of the dilute instanton-gas approximation \cite{coleman,*rajaraman}, we focus on the subset of paths satisfying these boundary conditions which correspond to a single instanton (i.e., those that pass over the barrier only once): $\vec\phi^{(1)}(\tau)$. Anticipating a semiclassical, stationary-phase approximation, we further subdivide these paths into two groups $\vec\phi^{(1)}_L(\tau)$ and $\vec\phi^{(1)}_R(\tau)$, corresponding to those suitably close to one or the other of the two classical, extremal paths $\vec\phi^\textrm{cl}_L(\tau)$ and $\vec\phi^\textrm{cl}_R(\tau)$:

\begin{eqnarray}
\langle -z|e^{-\frac{\hat{H}\tau}{\hbar}}|+z\rangle^{(1)}=\oint\mathcal{D}[\vec\phi_L^{(1)}(\tau)]e^{-\mathcal{S}_E[\vec\phi_L^{(1)}(\tau)]}\hspace{0.5cm}\nonumber\\
+\oint\mathcal{D}[\vec\phi_R^{(1)}(\tau)]e^{-\mathcal{S}_E[\vec\phi_R^{(1)}(\tau)]}\hspace{0.25cm}\label{eq:TLR}
\end{eqnarray}

Evaluating the functional integrals in eq.~\ref{eq:TLR} requires finding the classical extremal paths $\vec\phi^\textrm{cl}_L(\tau)$ and $\vec\phi^\textrm{cl}_R(\tau)$ which satisfy the instanton boundary conditions:

\begin{equation}
\vec\phi_T^\textrm{cl}(\tau)\rightarrow\begin{cases}
\left(-\phi_{Im},-\phi_{lm}\right)&\tau\rightarrow -\infty\\
\left(\phi_{Im},\phi_{lm}\right)&\tau\rightarrow\infty
\end{cases}\label{eq:bndy}
\end{equation}

\noindent where $T\in\{L,R\}$. Before tackling this, however, we note that there is already something we can say purely based on the boundary conditions of eq.~\ref{eq:bndy}. The Euclidean action associated with each of the two classical paths is composed of two terms:

\begin{equation}
\mathcal{S}_E[\vec\phi_T^{cl}(\tau)]\equiv\mathcal{S}_{T0}+\mathcal{S}_{Tb}\label{eq:action0}
\end{equation}

\noindent where the first is due to the tunneling dynamics, which we will discuss shortly, and the second to interaction with the bias source polarizing the island with charge $Q_b$. Due to the Josephson symmetry of the potential with respect to $\phi_I$, we can make the following statement about the latter term in the action:

\begin{eqnarray}
\mathcal{S}_{Rb}-\mathcal{S}_{Lb}&=&\frac{i}{\hbar}\int d\tau \vec{Q}_b\cdot\left[\frac{d\vec\Phi_R^\textrm{cl}(\tau)}{d\tau}-\frac{d\vec\Phi_L^\textrm{cl}(\tau)}{d\tau}\right]\nonumber\\
&=&\frac{i}{\hbar}\left[\oint_R Q_I d\Phi_I -\oint_L Q_I d\Phi_I\right]\nonumber\\
&=&\frac{i}{\hbar}\oint_I Q_I d\Phi_I\label{eq:loop}\\
&=& 2\pi i\times\frac{Q_b}{2e}\equiv iq_b\label{eq:phig}
\end{eqnarray}

\noindent where these contributions to the Euclidean action are imaginary because the voltage $\vec{V}=d\vec\Phi/dt\rightarrow -i\vec{V}$ under the Wick rotation to imaginary time $t\rightarrow i\tau$. The integration path subscripted $I$ in eq.~\ref{eq:loop} indicates a single, closed path encircling the island, which corresponds to a translation by exactly one period of the Josephson symmetry (note that this symmetry is not broken when $\delta\phi^x\neq0$). Therefore, as shown in eq.~\ref{eq:phig} and expected from our intuition for the Aharonov-Casher effect, the relative phase between transition amplitudes corresponding to the two classical tunneling pathways is given simply the dimensionless polarization charge applied to the island $q_b$. This relative phase shift can be viewed as a geometric effect \cite{berrypath}, associated with the area enclosed by the path $I$ in $(\Phi,Q)$ phase space \cite{MSPRA,AJKlongit},\footnote{Note that since $\phi_I$ can be viewed as an angular coordinate whose corresponding angular momentum is $q_I$, the geometric phase $q_b$ can be viewed as arising from a Sagnac-like effect, where the island charge offset corresponds to a constant angular ``rotation" of the system.}.

We now return to the problem of calculating the first term in eq.~\ref{eq:action0}, starting by writing down the two saddle-point energy barriers separating the persistent-current states, and traversed by the two tunneling paths $T\in\{L,R\}$:

\begin{eqnarray}
E_{bT}&\equiv&E_b\pm\delta E_b\nonumber\\
E_b&=&2\tilde{E}_{Ja}(\phi_\Delta)\tan\tfrac{\theta}{2}\nonumber\\
\delta E_b&=&\delta\phi^x\frac{E_b}{2}\left(\cot\theta+\csc\theta\right)\tan\tfrac{\phi_\Delta}{2}\label{eq:ETb}
\end{eqnarray}

\noindent where $\pm$ here and below refer to these two paths. From eqs.~\ref{eq:ETb} we see that when $\tilde{\delta\phi}^x\neq 0$, one of the barriers is lowered and the other is raised, as required to realize the situation pictured in fig.~\ref{fig:JPSQfig1}(d). Now, although the aforementioned extremal paths contain these points, solving exactly for these paths in 2+1D (already neglecting motion along the $\phi_p$ direction as discussed above) requires classically integrating the equations of motion, taking into account the fact that the inverse capacitance matrix has nondegenerate eigenvalues (that is, the fictitious phase particle has an anisotropic ``mass"). Since we are seeking here to obtain simple, analytic expressions useful for understanding the qualitative physics of this circuit, we use a simpler approach: for each group of paths $\vec\phi_T^{(1)}(\tau)$ ($T\in\{L,R\}$), we confine the functional integral to linear paths in ($\phi_l,\phi_I$) space connecting both potential minima and the intervening saddle point, as illustrated in fig.~\ref{fig:JPSQpot}(b)-(e). To facilitate this, we rotate to a set of coordinates $(\phi_T,\phi_T^\perp)$ for each path such that the approximate tunneling dynamics occurs only along the $\phi_T$ direction (and we can neglect the contributions from motion along $\phi_T^\perp$, by approximating it as separable and harmonic):

\begin{eqnarray}
\phi_T&\equiv&\left[\phi_I+\phi_l\frac{4\theta}{\pi}\left(1\pm\delta\phi^x\tan\tfrac{\phi_\Delta}{2}\frac{\cot\theta}{\pi}\right)\right]\frac{1}{N_+}\label{eq:phiT}\nonumber\\
q_T&\equiv&q_I\left[1\mp\delta\phi^x\tan\tfrac{\phi_\Delta}{2}\frac{32\theta^2\cot\theta}{\pi^3N_+}\right]\nonumber\\
&&\hspace{1cm}+q_l\frac{4\theta}{\pi}\left[1\pm\delta\phi^x\tan\tfrac{\phi_\Delta}{2}\frac{\cot\theta}{\pi}\frac{N_-}{N_+}\right]\nonumber\\
N_\pm&\equiv&1\pm\left(\frac{4\theta}{\pi}\right)^2\label{eq:coords}
\end{eqnarray}

\noindent where we have expanded to leading order in $\delta\phi^x$. We have also chosen the overall scaling of $\phi_T$ so that the sum of the lengths of the two 1D tunneling paths is $2\pi$, and the resulting problem can be viewed as occurring on a circle, as illustrated in fig.~\ref{fig:JPSQpot}(f).

Using eqs.~\ref{eq:phiT} and ~\ref{eq:JPSQpot}, we can evaluate the resulting 1D potentials along the two paths $U_T(\phi_T)\equiv U_J(\phi_T,\phi_T^\perp=0)$, examples of which are shown in figs.~\ref{fig:JPSQpot}(f). Along these paths, the potential minima are located at:

\begin{eqnarray}
\phi_{Tm}&\equiv&\phi_m\pm\delta\phi_m\nonumber\\
\phi_m&=&\frac{\pi}{2}\nonumber\\
\delta\phi_m&=&-\delta\phi^x\frac{\cot\theta}{2}\tan\tfrac{\phi_\Delta}{2}\frac{N_-}{N_+}\label{eq:phiTm}
\end{eqnarray}

\noindent and the effective Josephson inductances $L^{-1}_{Tm}\equiv L^{-1}_m+\delta L^{-1}_m$ at these minima are given by:

\begin{eqnarray}
L^{-1}_m&=&2\tilde{L}_{Ja}^{-1}\sin\theta\left[1+\tfrac{4\theta^2}{\pi^2\sin^2\theta}\right]\nonumber\\
\delta L^{-1}_m&=&-\delta\phi^x\tan\tfrac{\phi_\Delta}{2}\frac{4\theta \tilde{L}_{Ja}^{-1}}{\pi\sin\theta}\left[1+\tfrac{4\theta}{\pi^2\tan\theta}\left(1-\tfrac{2-8\sin^2\theta}{N_+}\right)\right]\nonumber\\
\tilde{L}_{Ja}^{-1}&\equiv&\tilde{E}_{Ja}\left(\frac{2\pi}{\Phi_0}\right)^2\label{eq:Linv}
\end{eqnarray}

\noindent We can make the corresponding transformation of the circuit's inverse capacitance matrix, starting from the $\{q_I,q_l,q_p\}$ representation, where it is given by:

\begin{eqnarray}
\begin{pmatrix}
      \frac{1}{C_I^\textrm{tot}} & 0 & \frac{C_{Ja}}{C_I^\textrm{tot}C_l^\textrm{tot}}\\

      0 & \frac{1}{C_l^\textrm{tot}} & 0 \\

      \frac{C_{Ja}}{C_I^\textrm{tot}C_l^\textrm{tot}} & 0 & \frac{1}{C_p^\textrm{tot}}
\end{pmatrix}\nonumber\\
\label{eq:Cinvmat}
\end{eqnarray}

\noindent with the definitions: $C_I^\textrm{tot}\equiv C_I+2\left(2C_{Ja}||C_{J}\right)$, $C_l^\textrm{tot}\equiv C_{Ja}+C_{J}/2$, and $C_p^\textrm{tot}\equiv 2C_{J}+\left(C_I||4C_{Ja}\right)$. From this, we obtain the inverse capacitance along the $q_T$ direction $C^{-1}_T\equiv C^{-1}+\delta C^{-1}$ \footnote{Note that this transformation also generates nonzero off-diagonal elements of the inverse capacitance matrix, corresponding to electrostatic coupling between the new charge variables $q_T$ and $q_T^\perp$. We neglect this coupling here on the grounds that the transverse motion has a much higher energy scale than the tunneling of interest here, so that their effect will be small.}:

\begin{eqnarray}
C^{-1}&=&\frac{1}{C_I}\left[\frac{\tfrac{1}{N_+^2}}{1+\tfrac{1}{r_C}\tfrac{1}{1+\tan\theta\sec\frac{\phi_\Delta}{2}}}+
\frac{\tfrac{16\theta^2}{\pi^2}\frac{4r_C}{N_+^2}}{1+\cot\theta\cos\frac{\phi_\Delta}{2}}\right]\nonumber\\
\delta C^{-1}&=&\delta\phi^x\tan\tfrac{\phi_\Delta}{2}\frac{1}{C_{Ja}}\left[\frac{\frac{16\theta^2}{\pi^2}\frac{2}{\pi N_p^2}}{\tan\theta+\cos\frac{\phi_\Delta}{2}}\right]\label{eq:Cinv}
\end{eqnarray}

\noindent where we have defined the ratio: $r_C\equiv C_I/4C_{Ja}$.

In order to find a simple analytic solution to the equations of motion in the two double-well potentials, we approximate them using the following sixth-order polynomial form:

\begin{equation}
U_T\approx E_{bT}\left[1-\left(\frac{\phi_T}{\phi_{Tm}}\right)^2\right]^2\left[1+k_{6T}\left(\frac{\phi_T}{\phi_{Tm}}\right)^2\right]\label{eq:poly}
\end{equation}

\noindent By construction, eqs.~\ref{eq:poly} have the same barrier height $E_{bT}$ and potential minima at $\phi_T=\pm\phi_{Tm}$ as the exact potential derived from eqs.~\ref{eq:JPSQpot} and ~\ref{eq:coords}. We can match the Josephson inductance at the local minima $\phi_{Tm}$ as well by choosing the (small) sixth order correction coefficient $k_{6T}$ to be:

\begin{eqnarray}
k_{6T}&\equiv&\frac{\Phi_{Tm}^2L^{-1}_{Tm}}{8E_{bT}}-1\label{eq:k6}
\end{eqnarray}

\noindent Figure~\ref{fig:JPSQpot}(g) illustrates these potentials for the left and right tunneling paths, where solid lines are the exact results, and dashed lines the approximation described by eq.~\ref{eq:k6} (note that they are nearly indistinguishable).

Using this form for the potential, we can readily find solutions for the imaginary-time equation of motion at zero energy, given by:

\begin{equation}
\frac{C_T}{2}\left[\frac{d\Phi_T^\textrm{cl}(\tau)}{d\tau}\right]^2-U_T[\Phi_T^\textrm{cl}(\tau)]=0\label{eq:EOM}
\end{equation}

\noindent with the boundary conditions $\phi_T(-\infty)=-\phi_{Tm}$ and $\phi_T(\infty)=\phi_{Tm}$ (equivalent to real-time dynamics in the inverted potential $-U_T(\phi_T)$ at zero energy). The simple polynomial form of eq.~\ref{eq:poly} allows eq.~\ref{eq:EOM} to be integrated to obtain:

\begin{equation}
\phi_T^{cl}(\tau)=\frac{\phi_{Tm}\tanh{\left[\frac{\Omega_T(\tau-\tau_0)}{2}\right]}}{\sqrt{1+k_{6T}\sech^2{\left[\frac{\Omega_T(\tau-\tau_0)}{2}\right]}}}\label{eq:instanton}
\end{equation}

\noindent where $\tau_0$ is an arbitrary position in imaginary time (known as the collective coordinate of the instanton), and $\Omega_T\equiv\sqrt{L^{-1}_{Tm}C^{-1}_T}$ is the frequency of small oscillations about the potential minima. Using this solution, we can evaluate the corresponding Euclidean action analytically by expanding in powers of the small parameter $k_{6T}$ and integrating:

\begin{eqnarray}
\mathcal{S}_{T0}&=&\frac{1}{\hbar}\int_{-\infty}^\infty d\tau\left(\frac{C_T}{2}\left[\frac{d\Phi_T^\textrm{cl}(\tau)}{d\tau}\right]^2+U_T[\Phi_T^\textrm{cl}(\tau)]\right)\nonumber\\
&=&\frac{C_T}{\hbar}\int_{-\infty}^\infty d\tau\left[\frac{d\Phi_T^\textrm{cl}(\tau)}{d\tau}\right]^2\nonumber\\
&\approx&\frac{2}{3}\frac{\Phi_{Tm}^2}{\hbar Z_T}\left[1-\frac{2}{5}k_{6T}+...\right]\label{eq:action}\\
\nonumber
\end{eqnarray}

\noindent where $Z_T=\sqrt{C_T^{-1}/L_{mT}^{-1}}$. Using eqs.~\ref{eq:phiTm},~\ref{eq:Linv},~\ref{eq:Cinv}, and ~\ref{eq:k6}, we can now obtain:

\begin{eqnarray}
\mathcal{S}_{T0}&\equiv&\mathcal{S}_0\left(1+\delta s\right)\nonumber\\
\mathcal{S}_0&=&\frac{16}{3}\frac{E_b}{\hbar\Omega}\left[1+\frac{3}{5}k_6\right]\nonumber\\
\delta s &\approx& \frac{\delta E_b}{E_b}-\frac{1}{2}\left[\frac{\delta L^{-1}_m}{L_m^{-1}}+\frac{\delta C^{-1}}{C^{-1}}\right]\label{eq:Sinfo}
\end{eqnarray}

\noindent where the two terms in the last line correspond to the $\delta\phi^x$-dependence of the barrier height and oscillation frequency, respectively, $\Omega\equiv1/\sqrt{L_mC}$ is the average single-well oscillation frequency, and we have neglected the small contribution to $\delta s$ from the $\delta\phi^x$-dependence of $k_{6T}$, by defining: $k_6\equiv(k_{6L}+k_{6R})/2$ [c.f., eq.~\ref{eq:k6}].

To get a more intuitive picture of the important parameter dependencies, we consider the large-$\beta$ (small-$\theta$) limit, obtaining to leading order in $1/\beta$:

\begin{eqnarray}
\Omega&\approx&\omega_J\sqrt{\frac{\cos\tfrac{\phi_\Delta}{2}}{\beta(1+r_C)}\left(\frac{\pi^2+4}{\pi^2}\right)}\nonumber\\
\mathcal{S}_0&\approx& y_{Ja}\tan\tfrac{\phi_\Delta}{2}\sqrt{\beta(r_C+1)}\left(\frac{3\pi^2+44}{15\sqrt{\pi^2+4}}\right)\nonumber\\
\delta s&\approx&\beta\tan\tfrac{\phi_\Delta}{2}\left(1-\frac{1}{\pi}+\frac{2\pi}{\pi^2+4}\right)\delta\phi^x\label{eq:Sparam}
\end{eqnarray}

\noindent where we have defined the bare junction plasma frequency $\omega_J\equiv1/\sqrt{L_{Ja}C_{Ja}}$ and the dimensionless DC SQUID admittance: $y_{Ja}\equiv R_Q/Z_{Ja}$, with $Z_{Ja}\equiv\sqrt{\tilde{L}_{Ja}/2C_{Ja}}$, and $R_Q=h/4e^2$ is the superconducting resistance quantum. The parameter $y_{Ja}$ describes the importance of quantum phase fluctuations across each DC SQUID, with the large-$y_{Ja}$ limit corresponding to semiclassical behavior (small phase fluctuations), and the tunneling action proportional to $y_{Ja}$. This quantity, in combination with $\beta$ and $r_C$, are the fundamental dimensionless parameters of the circuit in this simplified model.

In order to perform the path integrals in eq.~\ref{eq:TLR}, we generalize well-known results for the quartic potential \cite{coleman,*rajaraman}. In computing the usual fluctuation determinant describing Gaussian fluctuations about each stationary path, we treat the effect of the small sixth-order term in each potential using first-order perturbation theory. We account for the contributions of both stationary paths by viewing our quasi-1D problem as a particle on a circle with two potential minima [c.f., fig.~\ref{fig:JPSQpot}(f)]. We obtain, in the dilute instanton gas approximation, the following result for the Euclidean transition amplitude:

\begin{widetext}
\begin{equation}
\lim_{\tau\rightarrow\infty} \langle -z|e^{-\frac{\hat{H}\tau}{\hbar}}|+z\rangle=\sqrt\frac{\Omega}{\pi}\textrm{exp}\left[-\frac{\tau}{2}\left(\Omega-
\Omega_{ge}\left[\cos\tfrac{q_b}{2}-i\sin\tfrac{q_b}{2}\left(\mathcal{S}_0-\tfrac{1}{2}\right)\delta s\right]\right)\right]\label{eq:mat}
\end{equation}
\end{widetext}

\noindent where we have defined the fluxon tunnel splitting frequency:

\begin{equation}
\Omega_{ge}\equiv\Omega\sqrt{\frac{12\mathcal{S}_0}{\pi}\left(1+\frac{4}{5}k_6\right)}e^{-\mathcal{S}_0}\label{eq:tunnel}
\end{equation}

\noindent Combining eqs.~\ref{eq:IpJPSQ} and ~\ref{eq:mat}, we obtain the following effective Hamiltonian for the two lowest-energy states (up to an overall energy offset, and a static rotation around $z$), valid for small $\delta\phi^z$, $\delta\phi^x$, and arbitrary $q_b$:

\begin{eqnarray}
\hat{H}=-\hat\sigma^z\delta\phi^z\tilde{E}_{Ja}(\phi_\Delta)\cos\theta\hspace{4cm}\nonumber\\
\hspace{1cm}+\frac{\hbar\Omega_{ge}}{2}\left[\hat\sigma^y\cos\tfrac{q_b}{2}
-\hat\sigma^x\sin\tfrac{q_b}{2}\left(\mathcal{S}_0-\tfrac{1}{2}\right)\delta s\right]\hspace{1cm}\label{eq:TLS}
\end{eqnarray}

\noindent As expected, when $Q_b=e$ ($q_b=\pi$) and $\delta\phi^z=\delta\phi^x=0$ in eq.~\ref{eq:TLS}, we have $\hat{H}=0$ due to destructive Aharonov-Casher interference between the two tunneling paths in eq.~\ref{eq:TLR} \footnote{Note that in varying $\delta\phi^x$ around this zero emulated field point, the effective Hamiltonian can be proportional to either $-\hat\sigma^x$ or $+\hat\sigma^x$ (the latter of which is impossible to realize with a conventional flux qubit).}; this corresponds to the emulated zero field point around which we wish to operate.

\subsection{Dipole moments of the JPSQ}\label{sub:dipoles}

Focusing on the regime near this point, we formally re-write the Hamiltonian of eq.~\ref{eq:TLS} in terms of a Zeeman-like interaction between an effective vector dipole moment operator $\hat{\vec\mu}$, and an effective field $\vec{\mathcal{F}}$, as follows:

\begin{equation}
\hat{H}\equiv-\underbrace{\left(-\frac{d\hat{H}}{d\vec{\mathcal{F}}}\right)}_{\hat{\vec\mu}}\cdot\underbrace{\left(\delta\Phi^x,\delta Q^y,\delta\Phi^z\right)}_{\vec{\mathcal{F}}}\label{eq:mu}
\end{equation}

\noindent where we have defined $\delta Q^y\equiv Q_b-e$, and $\hat{\vec\mu}$ is given at zero field by:

\begin{eqnarray}
\hat{\vec\mu}&\equiv&\left(I^x\hat\sigma^x,V^y\hat\sigma^y,I^z\hat\sigma^z\right)\nonumber\\
I^x&=&e\Omega_{ge}\left(\mathcal{S}_0-\tfrac{1}{2}\right)\frac{\delta s}{\delta\phi^x}\nonumber\\
V^y&=&\frac{\Phi_0\Omega_{ge}}{4}\nonumber\\
I^z&=&\tilde{I}_{Ca}(\phi_\Delta)\cos\theta\label{eq:dipoles}
\end{eqnarray}

\noindent These dipole moments govern the strength with which the JPSQ couples (in the computational space) to external fields, including classical fields used to manipulate it, fields from other qubits that are used to engineer entangling interactions, and fields from its noise environment that are responsible for decoherence. This is, of course, a general feature of nearly any qubit system, that the same interactions with external fields that provide a mechanism for using the qubit also open the door to decoherence processes.

\begin{figure*}
    \begin{center}
    \includegraphics[width=1.0\linewidth]{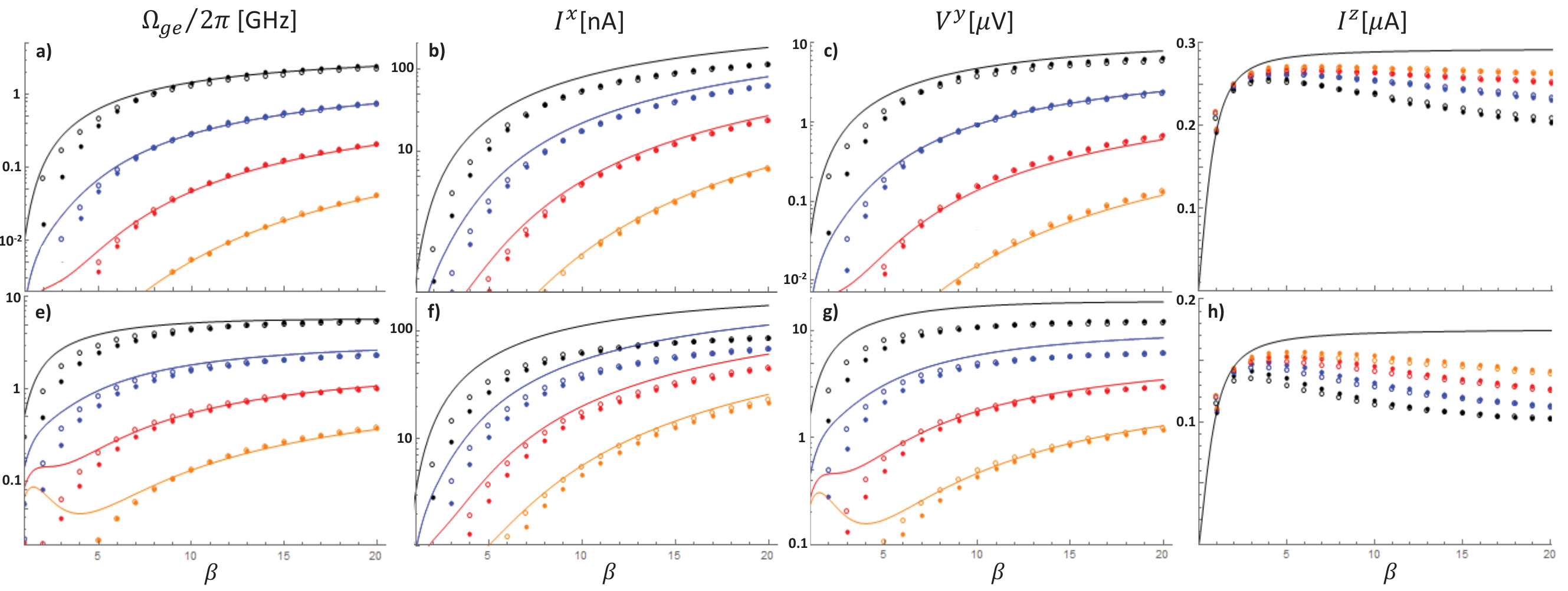}
    \caption[]
        {Comparison of predicted JPSQ energy splitting and dipole moments (solid lines) with full numerical simulations (symbols). Colors indicate different values for the island shunt capacitance ratio $r_C\equiv C_I/4C_{Ja}$: black-0.1, blue-1, red-2.5, orange-5. Panels (a)-(d) are for $y_{Ja}=25$ and (e)-(h) are for $y_{Ja}=15$. Filled symbols are simulations of the circuit discussed in detail in this section, and shown in fig.~\ref{fig:JPSQfig1}. Open symbols are simulations of the corresponding RF-SQUID-like circuit, with the larger Josephson junctions in fig.~\ref{fig:JPSQfig1} replaced by linear inductors equal to their corresponding Josephson inductances $L_{J}=(\Phi_0/2\pi)^2/E_{J}$. Other parameters for (a) are: $E_{Ja}=h\times125$ GHz, $C_{Ja}=5.0$ fF, and (b): $E_{Ja}=h\times74.6$ GHz, $C_{Ja}=3.0$ fF. The flux offset is $\Phi_\Delta=0.3\Phi_0$ in all cases.}
        \label{fig:splitdipplot}
    \end{center}
\end{figure*}

Figure ~\ref{fig:splitdipplot} shows a comparison between the predictions of eqs.~\ref{eq:TLS}-\ref{eq:dipoles} and full numerical simulations of the circuit, performed using a generalization of the methods described in refs. ~\onlinecite{MITLLflux,MITLLcoupled} \cite{JJcircuitSim}. The abcissa for the plots is the parameter $\beta(\phi_\Delta)$, which describes the ratio between the Josephson inductance of the DC SQUIDs and that of the large junctions. Panels (a)-(d) are for $y_J=25$ and panels (e)-(h) are for $y_J=15$. The different colors in each plot indicate different values for the dimensionless island shunt capacitance $r_C$. The leftmost column, panels (a) and (e), shows the energy splitting $\Omega_{ge}/2\pi$ in a more conventional flux-qubit-like regime, where $\delta\phi^z=\delta\phi^x=Q_b=0$ (corresponding to zero $z$ and $x$ fields, and a maximal $y$ field). The remaining three columns show the three components of the dipole moment, $I^x$, $V^y$, and $I^z$, near the qubit's emulated zero-field point. Both the energy splittings and the dipole moments decrease strongly with increasing $r_C$, a trend that can be understood from eqs.~\ref{eq:Cinv} and ~\ref{eq:Sparam}: in the $\beta\gg1 (\theta\ll1)$ regime of most interest here, the effective inverse capacitance $C_T^{-1}$ that acts as a ``mass" for fluxon tunneling is mostly controlled by $C_I$. Note that this is somewhat different from an ordinary flux qubit, where the corresponding tunneling ``mass" is controlled largely by the capacitance \textit{across} the (single) small junction (or DC SQUID in the case of a two-loop qubit \cite{mooij,MITtunable}).

We make two general remarks about the agreement between our analytic results and the full simulations shown in fig.~\ref{fig:splitdipplot}. First, the agreement is much better for the larger value of $y_{Ja}$ (top row) than for the smaller (bottom row). The agreement is also better for larger values of $r_C$. Both of these trends are to be expected, since both of these parameters control the validity of the semiclassical (stationary phase) approximation used to derive eq.~\ref{eq:TLS}; that is, larger $y_{Ja}$ and $r_C$ both result in smaller quantum phase fluctuations. Second, in many cases the agreement is substantially worse for small $\beta$. This is also to be expected, since as $\beta$ is decreased, the relative importance of quantum fluctuations along the $\phi_p$ direction increases, as can be deduced from eq.~\ref{eq:JPSQpot} by calculating the effective impedances for small oscillations about the potential minima along the three mode directions.

\subsection{Numerical simulation of realistic JPSQ circuits}\label{s:realistic}

\begin{figure}
    \begin{center}
    \includegraphics[width=0.9\linewidth]{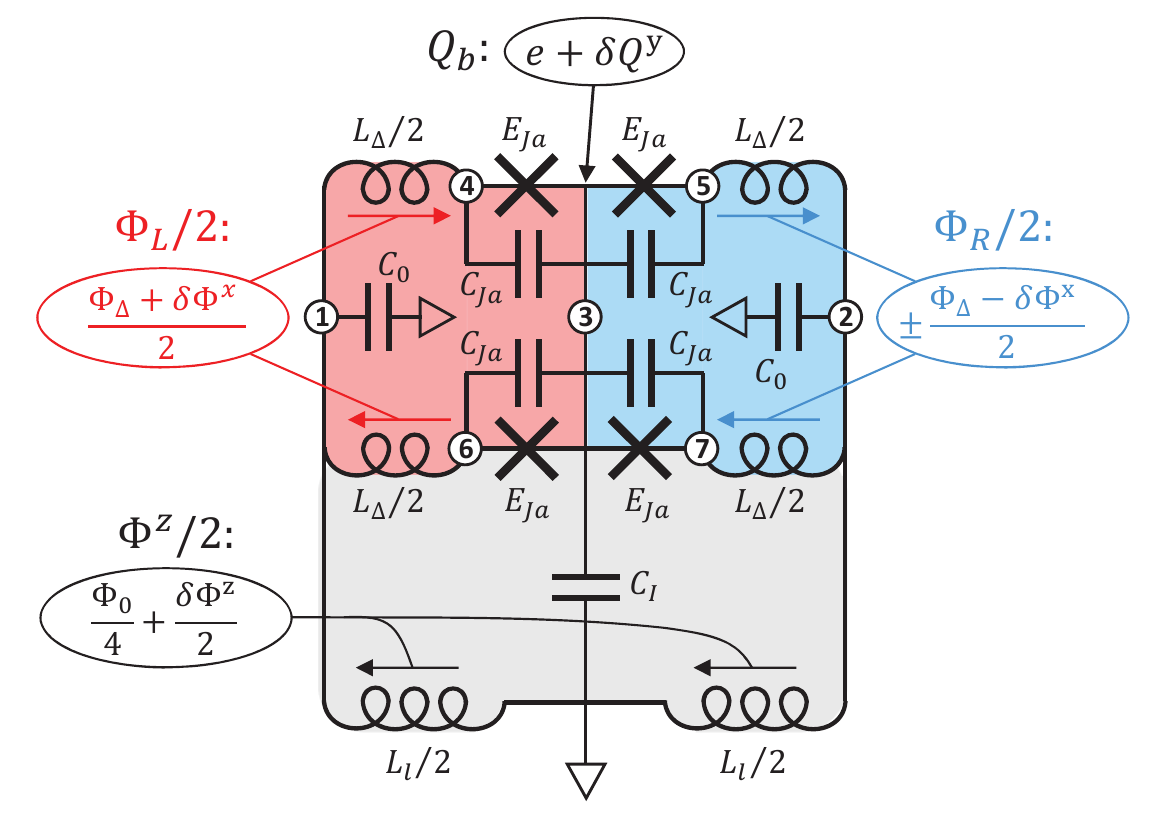}
    \caption[]
        {More detailed circuit model of JPSQ, including the linear inductances of both DC SQUID loops. For computational convenience, we have replaced the large Josephson junctions used in the circuit of fig.~\ref{fig:JPSQfig1} with linear inductors, in a manner analogous to that used in an RF-SQUID flux qubit. Colored arrows indicate flux bias offsets that are added to the gauge-invariant phase differences across the circuit's loop inductors. The circuit Hamiltonian is described in terms of one Josephson mode (described in a charge basis) and six oscillator modes (described in an oscillator eigenstate basis) \cite{JJcircuitSim}. Of the latter, three are linear oscillators and three have potentials that contain both Josephson and linear inductive terms. Note that the capacitors $C_0$ are required to make the capacitance matrix singular, but their numerical values are kept small enough to have negligible influence on the results.}
        \label{fig:JPSQfig2}
    \end{center}
\end{figure}

\begin{figure*}
    \begin{center}
    \includegraphics[width=1.0\linewidth]{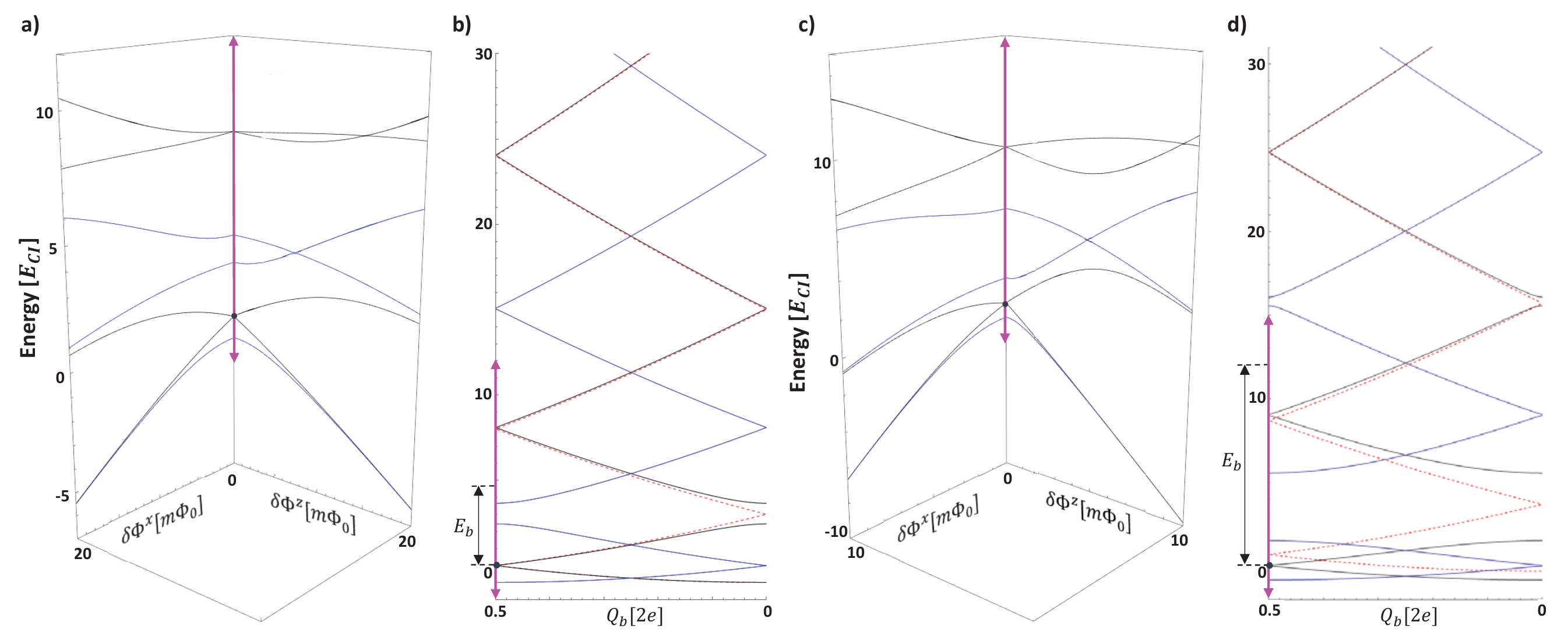}
    \caption[]
        {Numerical simulation of RF-SQUID JPSQ energy levels. Panels (a),(b) and (c),(d) correspond to the parameters of figs.~\ref{fig:splitdipplot}(a)-(d) and (e)-(h), respectively, for $\beta=15,r_C=0.1$, and with the loop inductances $L_l$ chosen to be equal to the corresponding total Josephson inductances [c.f.,~\ref{fig:JPSQfig1}(b)] for $\beta=15$: $72$ and $103$ pH, respectively. We have also taken $L_\Delta=50$pH, and $C_0=0.1$ fF. Note that $C_0$ must be nonzero for the capacitance matrix to be nonsingular, but the simulated energies depend negligibly on its numerical value as long at it does not become comparable to $C_{Ja}$. The two horizontal axes of (a) and (c) are the flux biases corresponding to the emulated $z$ and $x$ fields, and the zero field point is indicated by a filled black circle. Black (blue) solid lines are the eigenergies for even (odd) charge parity of the island. Panels (b) and (d) show the island charge dependence of the energy levels at this zero field point, where the vertical magenta arrows indicate how the levels between each pair of plots are connected: the island charge is varied from $Q_b=$0.5 to $Q_b=$0 starting from the zero field points in (a) and (c). A wider energy range is shown in (b) and (d) to illustrate the form of the island charge dependence: for energies greater than the top of the fluxon tunnel barrier (indicated  by $E_b$), the level structure increasingly matches that of the simple island charging Hamiltonian of eq.~\ref{eq:HC}. Red dashed lines in (b) and (d) show the eigenenergies of eq.~\ref{eq:HC}, where an overall energy offset has been added such that they match the simulation at high energy.}
        \label{fig:RFSQUIDplot}
    \end{center}
\end{figure*}

Although the circuit of fig.~\ref{fig:JPSQfig1}(b) and the corresponding results of eqs.~\ref{eq:TLS}-\ref{eq:dipoles} capture the most important qualitative features of the JPSQ, full numerical simulation and a more realistic circuit description are helpful to fill in additional important details. Figure ~\ref{fig:JPSQfig2} shows a JPSQ circuit which includes finite geometric loop inductances in the DC SQUIDs, and fig. ~\ref{fig:RFSQUIDplot} shows the low-lying energy levels obtained from numerical simulation \cite{JJcircuitSim,aboutJJ} of this circuit. The energies are plotted as a function of $\delta\Phi^z$, $\delta\Phi^x$, and $Q_b$, for parameters that correspond to those of fig.~\ref{fig:splitdipplot}, with $\beta=15,r_C=0.1$. In contrast to the analytic treatment in the previous section, numerical simulation allows us to look in detail at the higher energy levels of the circuit outside the computational subspace, which is important for understanding when these higher levels can safely be neglected. As shown in figs.~\ref{fig:RFSQUIDplot}(b) and (d), at energies greater than $\sim E_b$ (the height of the fluxon tunnel barrier, labelled in fig.~\ref{fig:RFSQUIDplot} with a vertical arrow) above the ground state, the $Q_b$-dependence increasingly looks like that of the simple charging Hamiltonian for the circuit's island, given by (up to a constant energy offset):

\begin{equation}
 \hat{H}_Q=4E_{CI}\left(\frac{\hat{Q}_I-Q_b}{2e}\right)^2\label{eq:HC}
\end{equation}

\noindent where $E_{CI}\equiv e^2/2C_I^\textrm{tot}$ is the island charging energy and $C_I^\textrm{tot}=4C_{Ja}+C_I$ for the circuit of fig.~\ref{fig:JPSQfig2}. Referring back to fig.~\ref{fig:JPSQpot}(a), we can immediately get an intuitive picture for the energy of the lowest excited states above the computational space: as the energy increases above the height of the fluxon tunnel barriers, the double-well potential in each unit cell along the $\phi_l$ direction becomes increasingly unimportant, and the states look more and more like plane waves (charge states) along the $\phi_I$ direction, governed by eq.~\ref{eq:HC}. This allows us to identify the next two higher excited states near $Q_b=e$ approximately with the two island charge states $\langle \hat{Q}_I\rangle/2e=\{-1,2\}$, whose energy above the qubit levels is $\approx8E_{CI}$. These levels will act as an upper bound on the energy scale over which we can treat the JPSQ as a two-level quantum system (similar to the so-called ``anharmonicity," which is used to specify the importance of the second excited state in the context of transmon and flux qubits).

Another piece of information evident in fig.~\ref{fig:RFSQUIDplot} is the importance of the island parity, indicated by black and blue lines for even and odd island parity states, respectively. Focusing on the region around $\delta\phi^z=0,Q_b=e$, we see that the odd-parity ground state is actually lower in energy than the two lowest even-parity states which form the computational subspace. This means that these even parity states can \textit{decay} into the lower-energy odd parity ground state near this point, if a quasiparticle tunnels onto the island inelastically \cite{qpcatelani}. Hence, the so-called ``parity lifetime" of the island, a quantity well-known in the literature of superconducting and semiconducting qubits \cite{qpcatelani,qpdevoret,parityDiCarlo,*paritymarcus,*parityLeo}, is crucially important in assessing the potential coherence of the JPSQ. Using the fact that the quasiparticle tunneling operator for a JJ which connects its even and odd-parity charge subspaces is proportional to $\sin{(\hat\phi/2)}$, where $\hat\phi$ is its gauge-invariant phase difference operator \cite{qpcatelani,qpdevoret}, one can readily verify that these inelastic, island-parity-changing quasiparticle tunneling events are strongly allowed in the JPSQ for all bias conditions \footnote{This stands in marked contrast to single-loop flux \cite{MITLLflux} or fluxonium \cite{manucharyan} qubits which have only one, single-JJ fluxon tunneling path \cite{qpdevoret}, and for which such processes are forbidden by symmetry near $\phi^z=\pi$. Qualitatively, this difference can be understood from the fact that for $\phi^z=\pi$, at the semiclassical peak of the tunnel barrier, the $\pi$ phase difference appears across a single JJ for the flux or fluxonium qubits, but across two series JJs (or SQUIDs) for the JPSQ, such that the operator $\sin{(\hat\phi/2)}$ has a well-defined, even symmetry in the former case, but not the latter. Note that for two-loop flux or fluxonium qubits, the symmetry which prevents inelastic quasiparticle tunneling at $\phi^z=\pi$ only exists when no flux is applied to the DC SQUID loop.}. As a result of this, both the ground and excited states in the (even parity) computational subspace will have decay rates $1/T_1$ of the order of the quasiparticle current spectral density $S_\textrm{qp}(\omega)$ defined in ref.~\onlinecite{qpcatelani}. For Aluminum junctions, typical quasiparticle densities, and parameters in the range considered here, these lifetimes could be as short as the $\sim\mu$s range. Fortunately, there exists a method for almost completely suppressing these processes, which has been used in both superconducting and semiconducting circuits to increase island parity lifetimes to the millisecond range and beyond \cite{parityDiCarlo,*paritymarcus,*parityLeo}. Referring to the circuit of fig.~\ref{fig:JPSQfig2}, one need only use a higher-gap superconducting material for the island as compared to the rest of the circuit, such that quasiparticles occupying the island see a higher potential energy. Once this potential barrier becomes substantially larger that the sum of the thermal energy and the maxmimum emulated Zeeman energy, the circuit is effectively protected from noise-induced parity-changing transitions.

\subsection{Discussion of JPSQ coherence and comparison with existing superconducting qubits}\label{sub:coherence}

\begin{table*}
\begin{threeparttable}[t]
\centering
\small
\begin{tabular}{c|ccc|cc|ccc|cc|ccc|cc}
\toprule[1pt]
JPSQ parameters &\multicolumn{5}{c|}{\textbf{A}} & \multicolumn{5}{c|}{\textbf{B}}& \multicolumn{5}{c}{\textbf{C}} \\
\hline
&&\multicolumn{2}{|c|}{numerical}&\multicolumn{2}{c|}{coherence}&&\multicolumn{2}{|c|}{numerical}&\multicolumn{2}{c|}{coherence}&&\multicolumn{2}{|c|}{numerical}&\multicolumn{2}{c}{coherence}\\
moment &eq.~\ref{eq:dipoles}&\multicolumn{2}{|c|}{simulation}&$\Gamma^g_{\phi e}$ & $T_1$&eq.~\ref{eq:dipoles}&\multicolumn{2}{|c|}{simulation}&$\Gamma^g_{\phi e}$ & $T_1$&eq.~\ref{eq:dipoles}&\multicolumn{2}{|c|}{simulation}&$\Gamma^g_{\phi e}$ & $T_1$\\
 &  & \multicolumn{1}{|c}{fig.~\ref{fig:JPSQfig1}} & fig.~\ref{fig:JPSQfig2} &[$10^6$rad/s] & \;[$\mu$s]\;& & \multicolumn{1}{|c}{fig.~\ref{fig:JPSQfig1}} & fig.~\ref{fig:JPSQfig2}&[$10^6$rad/s] &\; [$\mu$s]\; &  & \multicolumn{1}{|c}{fig.~\ref{fig:JPSQfig1}} & fig.~\ref{fig:JPSQfig2}&[$10^6$rad/s] & \;[$\mu$s]\; \\
\hline
$I^x$[nA]     &43  & 34  & 36  & 1.2 & 540 & 93  & 75  & 79 & 2.6 & 110 & 250  & 200 & 230 & 7.4 & 14\\
$V^y$[$\mu$V] &2.6 & 2.2 & 2.3 & 2.7 & 300 & 5.5  & 4.9 & 5.1 & 6.0 & 62 & 15   & 12  & 13 & 15 & 9.7\\
$I^z$[nA]     &69  & 41  & 40  & 1.3 & 430 & 150  & 96  & 91 & 3.0 & 84 & 410  & 250 & 240 & 7.7 & 13\\
\hline
parameters & \multicolumn{1}{c}{$E_\textrm{Ja}$} & \multicolumn{1}{c}{$C_\textrm{Ja}$}& \multicolumn{1}{c}{$C_\textrm{I}$} & \multicolumn{1}{c}{$y_J$} & \multicolumn{1}{c|}{$r_C$} & \multicolumn{1}{c}{$E_\textrm{Ja}$} & \multicolumn{1}{c}{$C_\textrm{Ja}$}& \multicolumn{1}{c}{$C_\textrm{I}$} & \multicolumn{1}{c}{$y_J$} & \multicolumn{1}{c|}{$r_C$} & \multicolumn{1}{c}{$E_\textrm{Ja}$} & \multicolumn{1}{c}{$C_\textrm{Ja}$}& \multicolumn{1}{c}{$C_\textrm{I}$} & \multicolumn{1}{c}{$y_J$} & \multicolumn{1}{c}{$r_C$} \\
& \multicolumn{1}{c}{[GHz]} & \multicolumn{1}{c}{[fF]}& \multicolumn{1}{c}{[fF]} & \multicolumn{1}{c}{} & \multicolumn{1}{c|}{} & \multicolumn{1}{c}{[GHz]} & \multicolumn{1}{c}{[fF]}& \multicolumn{1}{c}{[fF]} & \multicolumn{1}{c}{} & \multicolumn{1}{c|}{} & \multicolumn{1}{c}{[GHz]} & \multicolumn{1}{c}{[fF]}& \multicolumn{1}{c}{[fF]} & \multicolumn{1}{c}{} & \multicolumn{1}{c}{} \\
\hline
values &29.8  & 1.44  & \multicolumn{1}{c}{60} & 5.1 & 10 & 65.6  & 2.64  & \multicolumn{1}{c}{20} & 10 & 1.9 & 174  & 3.0 & \multicolumn{1}{c}{0} & 18 & 0 \\
\bottomrule
\end{tabular}
\caption{Dipole moments and coherence metrics for three JPSQ parameter cases. Cases A,B, and C are chosen to realize three different magnetic dipole moments: $\sim40$nA, $\sim80$nA, and $\sim250$nA, with $I^x\sim I^z$ in each case. We have fixed $\beta=15$ across all cases, with $\Phi_\Delta=0.3\Phi_0$. We tabulate numerical values for the dipole moments in each case, including for comparison the analytic results of eqs.~\ref{eq:dipoles}, and numerical simulations of the circuits in fig.~\ref{fig:JPSQfig1} and fig.~\ref{fig:JPSQfig2}. From the latter, we derive the two coherence metrics $\Gamma^g_{\phi e}$ and $T_1$, which describe the longitudinal (dephasing) and transverse (population transfer) components of the decoherence, respectively. The first of these, $\Gamma^g_{\phi e}$, is defined in refs.~\cite{ithier,nakamura}, and characterizes the effect low-frequency noise; it is the rate coefficient in the Gaussian (for 1/f noise) expression for the envelope decay of the system's response to a spin-echo pulse sequence. Although pulsed operation is not the focus of this work, we use this metric because it can be specified without requiring artificial cutoff parameters for the noise spectrum \cite{ithier}. This complication is a result of the singular character of the noise at low frequencies, which also produces the non-exponential decay of coherence. Because of this non-exponential character, $(\Gamma^g_{\phi e})^{-1}$ cannot meaningfully be viewed as, or compared to, the usual ``dephasing time" $T_2$. The second metric is $T_1$, the lifetime of the qubit excited state. Here, we tabulate the lifetime that would result from the expected noise power spectral density at 5 GHz (resulting from an applied perpendicular field): $S_Q(h\times $5 GHz$)\sim(1.1 \times10^{-8}e)^{2}$/Hz, and $S_\Phi(h\times $5 GHz$)\sim(4.3\times10^{-11}\Phi_0)^{2}$/Hz. These noise levels are derived from the results of ref.~\cite{MITLLflux}, which were themselves obtained using data from a large number of flux qubits of varying design and over a range of bias conditions.}
\label{tab:JPSQcoh}
\end{threeparttable}
\end{table*}
\begin{centering}
\begin{table*}
\begin{threeparttable}[b]
\small
\begin{tabular}{c|c|cc|c|cc|c|cc|c|cc}
\toprule[1pt]
\;\;Qubit type\;& \multicolumn{3}{c|}{\textbf{Two-loop flux} \cite{MITtunable}}& \multicolumn{3}{c|}{\;\;\;\textbf{Fluxmon} \cite{fluxmon}\;\;\;}& \multicolumn{3}{c|}{\;\;\;\textbf{D-Wave flux} \cite{Dwaveflux}\;\;\;}&\multicolumn{3}{c}{\;\;\;\textbf{Transmon} \cite{transmon,*DiCarlotransmon,*martinistransmon}\;\;\;} \\
\hline

&&\multicolumn{2}{c|}{coherence}&&\multicolumn{2}{c|}{coherence}&&\multicolumn{2}{c|}{coherence}&&\multicolumn{2}{c}{coherence}\\
moment &\;\;approx.\;\;&$\Gamma^g_{\phi e}$ & $T_1$&\;\;approx.\;\;&$\Gamma^g_{\phi e}$ & $T_1$&\;\;approx.\;\;&$\Gamma^g_{\phi e}$ & $T_1$&\;\;approx.\;\;&$\Gamma^g_{\phi e}$ & $T_1$\\
&dipole&[$10^6$rad/s] & \;[$\mu$s]\;&dipole&[$10^6$rad/s] &\; [$\mu$s]\; &dipole &[$10^6$rad/s] & \;[$\mu$s]\;& dipole &[$10^6$rad/s] & \;[$\mu$s]\; \\
\hline
$I^x$[nA]      & 40\tnote{$\star$}  & 1.3 &\tnote{$\dagger$}& 40\tnote{$\star$}  & 1.3 & \tnote{$\dagger$} & 70\tnote{$\star$} & 7.4 & \tnote{$\dagger$}& N/A       &            &   \\
$V^y$[$\mu$V]  & 7.6\tnote{$\star$} & $\sim$0\tnote{$\ddag$}   & 28      & 3.4\tnote{$\star$} & $\sim$0\tnote{$\ddag$}   & 140       & 2.8\tnote{$\star$}& $\sim$0\tnote{$\ddag$}& 210 & 5 & $\sim$0\tnote{$\ddag$} & 62 \\
$I^z$[nA]      & 260         & 8.5 & 10      & 700         & 23  & 1.4       & 2600       & 82  & 0.10 &5\tnote{$\star$}&0.16\tnote{$\star$} & $\sim$0\tnote{$\dagger\dagger$}\;  \\
\bottomrule
\end{tabular}
\small
\begin{tablenotes}
\item [$\star$] Flux qubits' $I^x$ and $V^y$ go exponentially to zero as $\mathcal{F}^x\rightarrow 0$, while for a transmon $I^z$ goes linearly to zero near its maximum energy splitting (often called a ``sweet spot"). Here, we tabulate values for bias far from these points.
\item [$\dagger$] Since for flux qubits $I^x\rightarrow 0$ as $\mathcal{F}^x\rightarrow 0$, noise in $\mathcal{F}^x$ only produces $T_1$ processes when both $\mathcal{F}^x$ \textit{and} $\mathcal{F}^z$ are nonzero, and the resulting rate depends in detail on both of these quantities.
\item [$\ddag$] The very small static charge dispersion of these qubits makes them highly insensitive to dephasing from charge noise.
\item [$\dagger\dagger$] Viewed as a spin, the transmon can be described as experiencing a large offset field, which points purely along $z$ if its two junctions are symmetric. Therefore, flux noise in $\mathcal{F}^z$ can only produce nonzero transverse $T_1$ processes if this symmetry is broken.
\end{tablenotes}
\caption{Dipole moments and coherence metrics for demonstrated superconducting qubits. The first three columns are examples of persistent-current flux qubits, in order of increasing persistent current, with parameters taken from the indicated references (fluxonium \cite{manucharyan} is also a persistent current qubit; however, we have not included it in the present comparison because its persistent current is, by design, too small to be used for the direct spin emulation discussed here). The final column is for the transmon qubit widely used in gate-model applications, which is included as a point of reference.}
\label{tab:coh}
\end{threeparttable}
\end{table*}
\end{centering}

Assuming that the parity switching discussed above can be circumvented as in previous works \cite{parityDiCarlo,*paritymarcus,*parityLeo}, the dominant intrinsic sources of noise in these circuits will be the same as for other superconducting qubits. These can be described in terms of charge and flux noise, both of which exhibit high-frequency and ``1/f-like" low-frequency components that have been observed under a variety of conditions \cite{chargenoisekuzmin,*chargenoisemooij,*chargenoisemartinis,ithier,nakamura,bylander,MITLLflux,fluxmon}. As described above, the sensitivity of the JPSQ to charge and flux noise can be described simply in terms of its dipole moments [c.f., eqs.~\ref{eq:dipoles}]. Table ~\ref{tab:JPSQcoh} shows these moments calculated for three sets of JPSQ parameters, labelled cases A,B, and C, corresponding to circuit designs with increasing magnetic dipole moments (with $I^x\sim I^z$ in each case) spanning a range from $\sim$40 nA to $\sim$250 nA. Because the JPSQ is engineered to emulate a vector spin-1/2, these moments are, by design, approximately independent of field around the emulated zero-field point, allowing decoherence processes to be viewed in the same simple manner used to describe the response of a spin-1/2 to field noise. In the presence of a nonzero externally-applied offset field, noise fluctuations are naturally divided into their longitudinal and transverse components, relative to the axis of that field. Longitudinal noise fields cause the spin's Larmor precession frequency (Zeeman energy) to fluctuate, resulting in so-called ``dephasing" ($T_2$ processes) whose magnitude depends mostly on the low-frequency content of the noise power spectrum. These processes are represented by the quantity $\Gamma^g_{\phi e}$ shown in the table for each case, and for each moment. Transverse noise fields cause the spin's precession axis to rotate, resulting in population transfer ($T_1$ processes) in the computational basis that depends mostly on the noise power spectral density at the Larmor frequency, described by the quantity $T_1$ in the table.

For comparison, table ~\ref{tab:coh} provides typical values for existing, state-of-the art superconducting qubits. Its first three columns show values for three demonstrated examples of tunable flux qubits, to which the JPSQ is most sensibly compared, and its final column the well-known transmon qubit used for nearly all gate model applications \cite{transmon,*DiCarlotransmon,*martinistransmon}. Broadly speaking, tables ~\ref{tab:JPSQcoh} and ~\ref{tab:coh} exemplify the fact that unlike the transmon qubit, whose simplicity results in very little design freedom, the dipole moments of JPSQs and flux qubits can vary widely, according to the needs of the designer: tunable flux qubits have been used with $I^z$ values ranging from the $\sim50$nA in recent capacitively-shunted flux qubits \cite{MITLLflux} (with coherence times as high as $\sim50\mu$s), to $\sim3\mu$A in the machines from D-Wave systems \cite{Dwaveflux,Dwavecoupler,Dwave3DTFIM,*DwaveKT} (with coherence times in the range of tens of nanoseconds). Correspondingly, unlike the transmon qubit, whose coherence in the ideal case is largely set by fundamental material properties and physical geometry, the coherence of flux qubits and JPSQs also depends very strongly on the specific design requirements set by the system in which it is used \footnote{For example, the very large magnetic moments of the D-Wave flux qubits, which are responsible for their low coherence, are required to produce the very strong pairwise interactions needed for that system.}.  In spite of this fundamental difference, one simple coherence comparison that can still readily be made between the JPSQ and both flux and transmon qubits is in their sensitivity to high-frequency noise: the data for JPSQ cases A and B in table \ref{tab:JPSQcoh} clearly show that JPSQs can readily be designed with comparable or longer $T_1$ times than the flux qubits and transmon listed in table \ref{tab:coh} \footnote{Since both flux qubits and the JPSQ have more than one partial $T_1$, the full $T_1$, given an assumed direction of applied field, can be obtained simply from the parallel sum of the rates for the two moments perpendicular to this field. For example, for JPSQ case A in an applied $z$ field, one obtains: $T_1=540||300=190\mu$s.}.

In order to compare the dephasing in these circuits, we first note the following: both flux and transmon qubits have a so-called ``sweet spot," a bias point where their energy splitting becomes linearly insensitive to both low-frequency flux and charge noise (the latter is due to a large ratio of Josephson to charging energy, and is not dependent on bias). At such a bias point, we have, by definition: $\langle e|\hat{\vec\mu}|e\rangle-\langle g|\hat{\vec\mu}|g\rangle=0$; that is, at a sweet spot the qubit's static dipole moment is zero. This is obviously a useful property in some cases, as it allows the qubit to be decoupled from noise \cite{ithier,nakamura}; however, \textit{it is incompatible with emulation of a vector spin-1/2} (note that this is simply a more general restatement of the conclusion of section ~\ref{s:spinmodel}). Therefore, the JPSQ, or any circuit engineered to emulate a vector spin-1/2, must by its very design be open to additional dephasing channels as compared to flux and transmon qubits, which cannot. For the JPSQ, this additional dephasing manifests itself it two ways: First, compared to a conventional flux qubit, the JPSQ experiences dephasing due to noise in $\mathcal{F}^x$ at all values of $\Delta E^x$, whereas the flux qubit has a sweet spot when $\Delta E^x$ is at its minimum value, and is only subject to dephasing from $\mathcal{F}^x$ noise away from this point. As can be seen from the tables, the values for $I^x$ are generally comparable to or smaller than $I^z$; so, while this additional sensitivity will result in larger total dephasing rate, the difference will be less than a factor of two.

\begin{figure*}
    \begin{center}
    \includegraphics[width=1.0\linewidth]{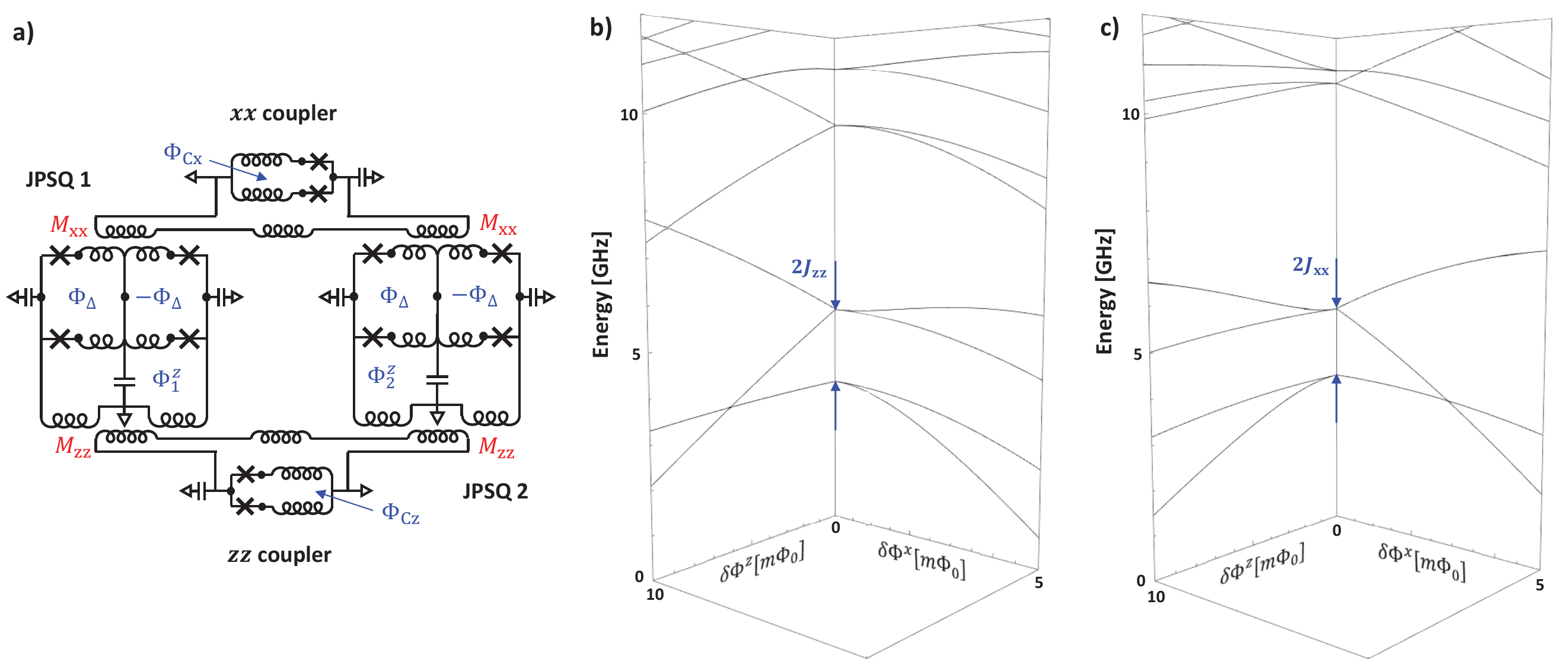}
    \caption[]
        {Simulated energy levels for two JPSQs coupled with a vector spin interaction in the $x-z$ plane. Panel (a) shows a schematic of the circuit, which contains two RF-SQUID-style JPSQs, each like that shown in fig.~\ref{fig:JPSQfig2}, with parameters: $E_{Ja}=h\times70.0$ GHz, $C_{Ja}=2.64$ fF, $L_l=1$ nH, $L_\Delta=50$ pH, $C_I=10$ fF, and $\Phi_\Delta=0.4\Phi_0$. These two qubits are coupled together by two, two-loop RF SQUID flux qubit couplers, with parameters: $E_J=h\times172$ GHz, $C_J=6.46$ fF, $L_l=600$ pH, $L_\Delta=10$ pH. The lower of these, labelled ``$zz$ coupler" in the schematic, is coupled via mutual inductances $M_{zz}=25$ pH to the loop inductors $L_l$ of the two JPSQs. The upper one, labelled ``$xx$ coupler", is coupled via mutual inductances $M_{xx}=25$ pH to both pairs of DC SQUIDs in the JPSQs. Because the two DC SQUIDs within each JPSQ are biased with opposite offset fluxes $\pm\Phi_\Delta$, the individual susceptibilities of the two fluxon tunneling amplitudes within each JPSQ are equal and opposite; therefore, the common mutual coupling $M_{xx}$ shown couples oppositely to these two amplitudes [c.f., fig.~\ref{fig:JPSQfig1}(d)], resulting in the desired $xx$ coupling. Panels (b) and (c) show the resulting energy levels when one of the couplers is turned on, and the other off, producing pure $zz$ coupling and pure $xx$ coupling, respectively. These two bias configurations show an evident symmetry under permutation of the labels $x$ and $z$ (Note that for clarity these plots show only the energies on the boundaries $(\delta\Phi^x,0)$ and  $(0,\delta\Phi^z)$ rather than full energy surfaces). This indicates that a true rotational symmetry (in this case around $y$) can be engineered, something that is not possible with any present-day superconducting qubit circuits.}
        \label{fig:twoJPSQ}
    \end{center}
\end{figure*}

Secondly, unlike both flux qubits and transmons, the JPSQ has a static $V^y$, making it sensitive to low-frequency charge noise, which is in general a more serious concern. Obviously, one would like therefore to minimize this sensitivity by keeping $V^y$ as small as possible. However, from eqs.~\ref{eq:dipoles} and ~\ref{eq:Sparam} we can see that the value of $V^y$ is closely tied to that of $I^x$. In fact, the exponential dependence of $I^x$ on the tunneling action $\mathcal{S}_0$ implies, for a given $I^x$ (assumed to be set by external system requirements), that $\beta\tan(\phi_\Delta/2)$ is the only accessible parameter on which the resulting $V^y$ depends more strongly than logarithmically (for $\beta\gg1$). This is evident in table \ref{tab:JPSQcoh}, where the ratio $V^y/I^x$ (with units of impedance) is nearly identical in all three cases, and leads to the same design conclusion as fig.~\ref{fig:fluxcomp}: that $\beta$ should be made as large as possible \footnote{In practical cases the maximum permissible size of $\beta$ will limited by the need to couple other circuits inductively to the qubit loop, and/or the maximum possible junction size if Josephson loop inductance is used.}. Examining the corresponding JPSQ dephasing rates due to low-frequency charge noise in table ~\ref{tab:JPSQcoh}, we see correspondingly that they are in all three cases about twice as large as those due to flux noise coupling to $I^x$. So, although this additional charge noise dephasing is unavoidable for the JPSQ, it will at worst increase the total dephasing rate only by this modest factor.

\section{Examples of multi-JPSQ circuits}\label{s:circuits}

We now give some examples of how spins emulated using the JPSQ circuit can be coupled together in ways that have not previously been possible with engineered quantum devices. Figure ~\ref{fig:twoJPSQ} shows our first example, in which two of the JPSQ circuits detailed in fig.~\ref{fig:JPSQfig2} are coupled to each other via a pair of tunable RF SQUID flux qubit couplers ~\cite{Dwavecoupler,MITLLcoupled}. One of these is magnetically coupled to the $I^z$ dipole moments of the qubits, and the other to their $I^x$ moments. Panels (b)and (c) show the lowest energy levels of this circuit, obtained by full numerical simulation ~\cite{JJcircuitSim,aboutJJ}. For panel (b), the $zz$ coupler is turned on ($\Phi_{Cz}=\Phi_0$) and the $xx$ coupler is off ($\Phi_{Cx}=\Phi_0/2$), while for panel (c) the reverse is true ($\Phi_{Cz}=\Phi_0/2,\Phi_{Cx}=\Phi_0$). Focusing on the lowest four energy levels, which act as the two-spin computational subspace, we see that strong two-qubit $zz$ and $xx$ coupling are both possible with this circuit just by adjusting the flux controls of the two couplers. Furthermore, the two types of coupling are qualitatively equivalent, as evidenced by the fact that the two panels, one with only $xx$ coupling turned on and the other with only $zz$, look nearly identical, except that the roles of $x$ and $z$ are permuted [compare to the strong non-equivalence seen for a two-loop flux qubit shown in fig.~\ref{fig:fluxqubitfig}(h)]. This equivalence is a result of the JPSQ's ability to simultaneously emulate a true rotational symmetry at its zero-effective-field point while maintaining a strong, vector dipole moment.

\begin{figure}
    \begin{center}
    \includegraphics[width=0.8\linewidth]{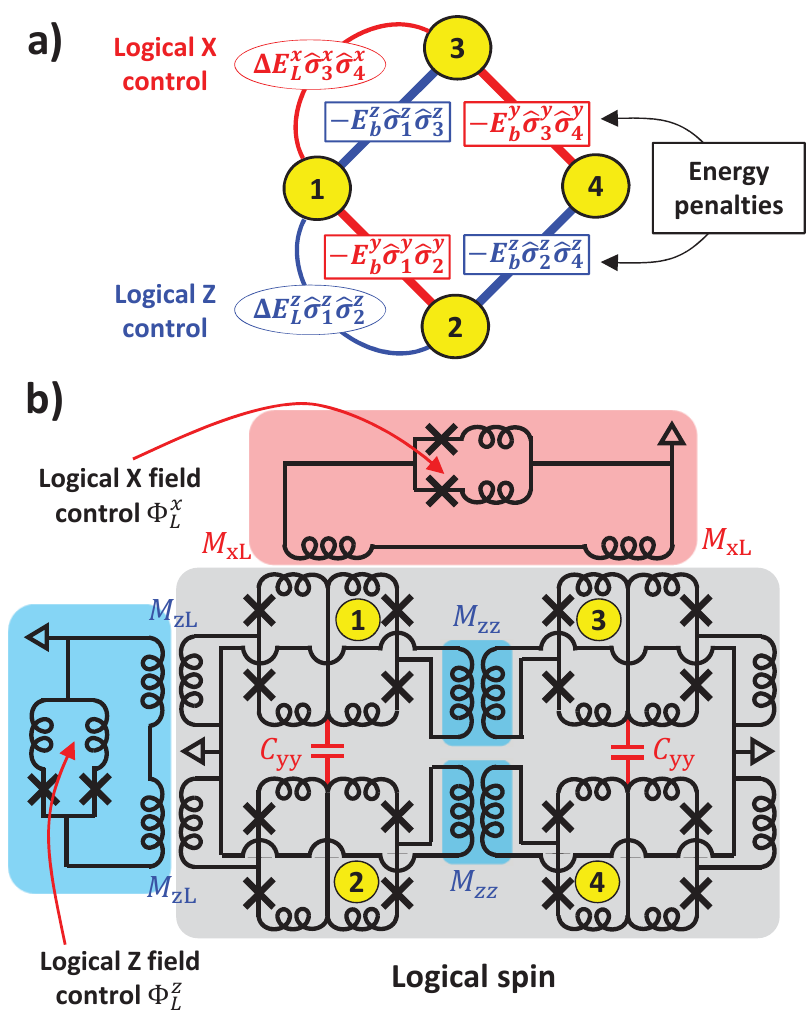}
    \caption[]
        {Passive quantum error suppression circuit based on a distance-2 Bacon-Shor code, using the JPSQ. Panel (a) illustrates the desired construction, consisting of four physical qubits. Strong, pairwise, static interactions corresponding to four commuting check operators of the code are used to shift up to higher energy all states for which any check operators have a positive eigenvalue. This creates an energy barrier for any physical processes that act locally on single qubits \cite{jiangrieffel,marvianlidar}. Weaker, two-qubit interactions are then used to realize the logical spin operators. Panel (b) shows a schematic of the circuit simulated here based on four JPSQs. The strong penalty interactions are realized with the static magnetic and electric couplings: $M_{zz}$ and $C_{yy}$. Tunable logical fields are realized using additional, two-loop, RF SQUID flux qubit couplers.}
        \label{fig:BSqubit}
    \end{center}
\end{figure}

\begin{figure*}
    \begin{center}
    \includegraphics[width=1.0\linewidth]{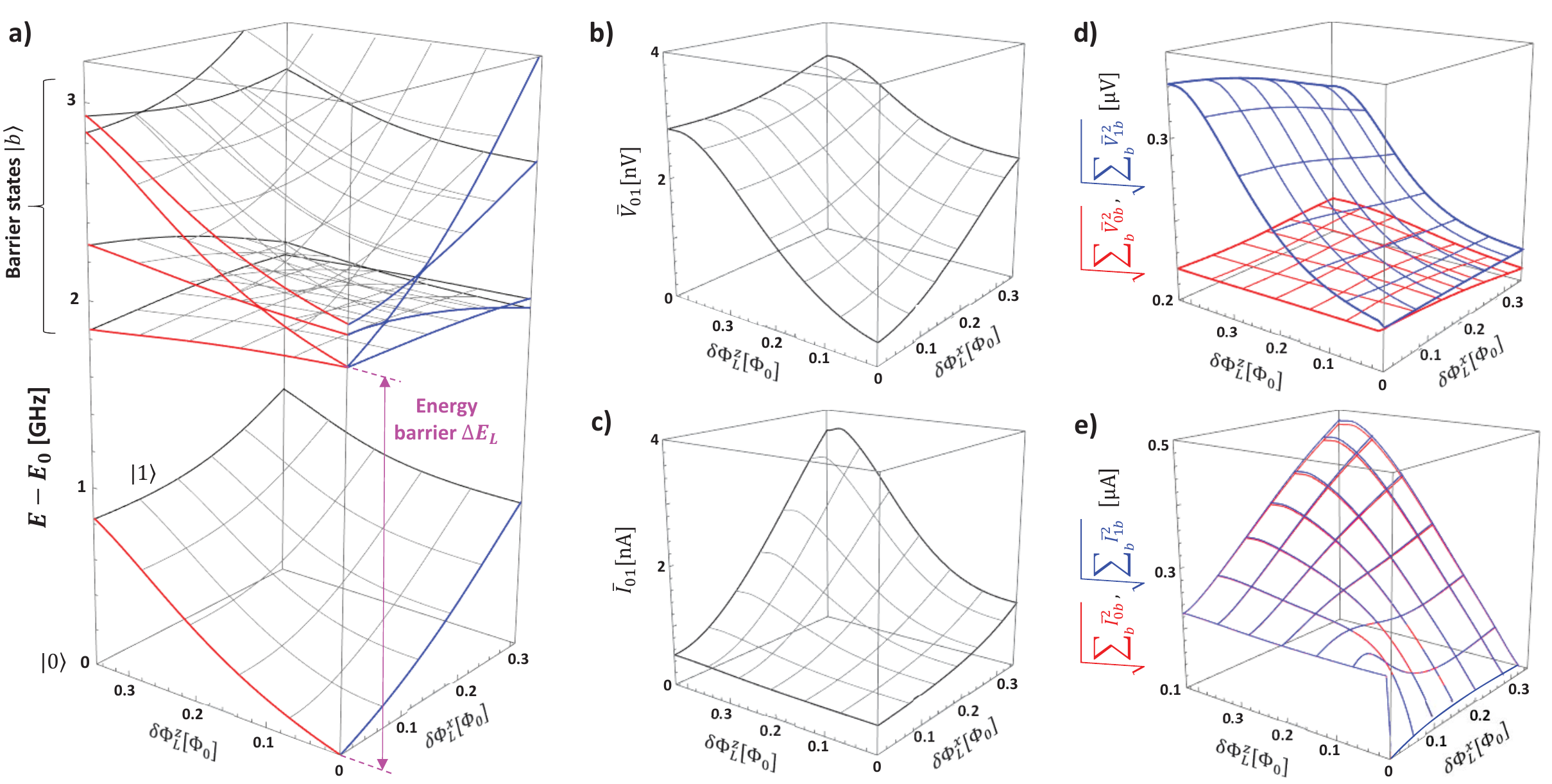}
    \caption[]
        {Simulation of the passive quantum error suppression circuit of fig.~\ref{fig:BSqubit}. Numerical parameter values used here for the JPSQs are: $E_{Ja}=h\times159$ GHz, $C_{Ja}=0.96$ fF, $C_I=1.0$ fF, $L_l=850$ pH, $L_\Delta=100$ pH, $\Phi_\Delta=0.4\Phi_0$. Note that smaller capacitances and larger $E_{Ja}$ are used here, as compared to table ~\ref{tab:JPSQcoh}, to increase the JPSQs' $I^z$ and $V^y$ dipole moments and the resulting energy barriers due to the $zz$ and $yy$ penalty interactions (with parameters $M_{zz}=110$ pH and $C_{yy}=30$ fF). For the RF SQUID couplers, we used: $E_J=h\times497$ GHz, $C_J=6$ fF, $L_l=120$ pH, $L_\Delta=10$ pH, $M_{zL}=M_{xL}=20$ pH. These values were chosen to minimize the flux noise sensitivity induced by the logical field control, for the pictured $h\times\sim1$ GHz maximum logical Zeeman splitting. Panel (a) shows the low-lying energy levels for the circuit (relative to the ground state energy), obtained using the methods of ref.~\onlinecite{JJcircuitSim,aboutJJ}, as a function of the two logical field control fluxes. The purple arrow indicates the energy gap between the two computational logical states and the lowest manifold of excited states which violate one of the penalty interactions. Panels (b) and (c) are plots of the total average electric and magnetic dipole moments [c.f., eq.~\ref{eq:Ldipoles}] of the logical states, with respect to physical noise fields, as a function of the logical $z$ and $x$ control fluxes (note that decoherence rates will in general scale with the \textit{square} of these quantities). Panels (d) and (e) give the total rms electric and magnetic dipole moments, respectively, for excitation out of the logical space (from logical state $|0\rangle$ shown in red, and from logical state $|1\rangle$ in blue) into the first excited manifold of states $|b\rangle$ shown in (a), separated by the energy barrier shown with a magenta arrow.}
        \label{fig:BSqubitplots}
    \end{center}
\end{figure*}

The circuits we have discussed so far exploited only the $x$ and $z$ components of the JPSQ dipole, which are both magnetic. Figure~\ref{fig:BSqubit} shows a circuit which makes use of all three of its vector components. This circuit contains four coupled JPSQs (each like the one shown in fig.~\ref{fig:JPSQfig2}) which together realize a single logical spin with passive quantum error suppression, based on a distance-2 Bacon-Shor code \cite{jiangrieffel,marvianlidar}. The JPSQs are connected by strong, pairwise couplings, with magnetic $zz$ couplings between the qubit pairs (1,3) and (2,4), and electric $yy$ couplings between pairs (1,2) and (3,4). These interactions correspond to the two-qubit check operators for the code, and together they produce a 2-dimensional logical computational subspace whose effective logical spin operators are products of two single-qubit physical operators, as shown in fig.~\ref{fig:BSqubit}(a). Because the logical spin operators are 2-local, the two logical states in the ideal case are protected against all single-qubit noise on the four constituent physical qubits; or, more precisely, any single qubit noise process should couple the two logical states only to higher-energy levels, separated by an energy ``barrier" whose height is of the order of the strength of the strong pairwise interactions. Therefore, single-qubit noise processes in the four constituent qubits in the ideal case must supply at least this amount of energy to affect the encoded logical spin state.

The circuit of fig.~\ref{fig:BSqubit}(b) also contains two additional two-loop RF SQUID flux qubit couplers (each like the couplers shown in fig.~\ref{fig:twoJPSQ}), which are used to implement a $zz$ interaction between qubits 1 and 3, and an $xx$ interaction between qubits 1 and 2. As illustrated in fig.~\ref{fig:BSqubit}, these two interactions correspond to the logical $x$ and $z$ operators, respectively. Adjusting the strength of these two-qubit couplings controls the effective field seen by the logical spin. Figure ~\ref{fig:BSqubitplots}(a) shows the resulting dependence of the simulated \cite{JJcircuitSim,aboutJJ} energy levels (relative to the ground state energy) on these two logical field controls. Around zero field, the resulting energy barrier separating the logical states from the first excited manifold (corresponding to violation of one of the penalty interactions) is $\sim h\times2.1$ GHz$\sim5.3k_BT$, corresponding to a Boltzmann factor of $5\times10^{-3}$. This number can be viewed as the relative thermal occupation of environmental photons at the low frequencies which separate the two logical states, and those that connect them both to the next higher set of states.

We now wish to evaluate the sensitivity of this protected qubit to decoherence arising from local, physical flux and charge noise. We define the c-number dipole moment  $\mathcal{D}_{ij}[\hat{\mathcal{O}}]$ associated with a dipole operator $\hat{\mathcal{O}}$ and a two-dimensional subspace $\{|i\rangle,|j\rangle\}$ as:

\begin{equation}
\mathcal{D}_{ij}[\hat{\mathcal{O}}]\equiv\frac{1}{2}\mathcal{W}\left[\mathcal{O}_{ss^\prime}\right]\label{eq:dipoleop}
\end{equation}

\noindent where $\mathcal{W}\left[\mathcal{O}_{ss^\prime}\right]$ is the numerical range of the $2\times2$ matrix $\langle s|\hat{\mathcal{O}}|s^\prime\rangle$, and $s,s^\prime\in\{i,j\}$. We make the simplifying assumption that each circuit node experiences independent charge noise of the same magnitude, and each geometric inductor experiences independent flux noise of the same magnitude. We can then define the following effective total electric and magnetic dipole moments to which these common noise levels can be said to couple in the two-dimensional subspace $\{i,j\}$:

\begin{eqnarray}
\bar{V}_{ij}&\equiv&\sqrt{\sum_{n\in\textrm{nodes}}\left(\mathcal{D}_{ij}\left[\hat{V}_n\right]\right)^2}\nonumber\\
\bar{I}_{ij}&\equiv&\sqrt{\sum_{l\in\textrm{inductors}}\left(\mathcal{D}_{ij}\left[\hat{I}_{l}\right]\right)^2}\label{eq:Ldipoles}
\end{eqnarray}

\noindent and which can be compared to the corresponding dipole moments of physical qubits listed in tables ~\ref{tab:JPSQcoh} and ~\ref{tab:coh}. Figure ~\ref{fig:BSqubitplots}(b) and (c) show the results for these dipole moments in the logical computational subspace ($i,j=1,2$) derived from our simulations of the circuit of fig.~\ref{fig:BSqubit}(b). The electric dipole moment is in the $\sim$nV range, \textit{thousands} of times smaller than typical superconducting qubits, suggesting that its $T_1$ due to charge noise would be \textit{millions} of times longer \footnote{Note that for the ideal case illustrated in fig.~\ref{fig:BSqubit}(a), this electric dipole would be identically zero. However, for the actual circuit of fig.~\ref{fig:BSqubit}(b), additional effective interactions (mediated by higher excited states) allow local charge noise to couple to two-qubit operators in the logical computational subspace.}. The magnetic dipole moments shown in fig. ~\ref{fig:BSqubitplots}(c) are in the $\sim$nA range, set completely by the fact that we have intentionally engineered logical field controls via the two additional couplers [c.f., the shaded red and blue subcircuits in fig.~\ref{fig:BSqubit}(b)] and in so doing to we have effectively ``poked a hole" in the error suppression. The $\sim$nA scale of these magnetic dipoles results simply from choosing to produce an $\sim h\times1$ GHz logical Zeeman splitting over the maximum coupler control flux tuning range of $\Phi_0/2$. Much smaller magnetic dipole moments could trivially be achieved simply by reducing the accessible tuning range of the Zeeman energy; in the extreme case where we remove the logical field control couplers entirely (reducing this tuning range to zero), our simulations show residual magnetic dipole moments less than 1 pA. The key question here, in assessing the potential coherence of such a logical spin, is the manner in which it would be used. In the context of quantum annealing, one would like to retain a wide tuning range for the Zeeman energy, and the $\sim$ nA dipole moments shown in fig.~\ref{fig:BSqubit}(e) are therefore likely the smallest possible, if a single loop is used for logical field control (restricting the full flux tuning range to $\le\Phi_0/2$). However, this is already $\sim$100-10,000 times smaller than the magnetic dipoles of existing flux qubits [c.f., table~\ref{tab:coh}].

\begin{figure*}
    \begin{center}
    \includegraphics[width=1.0\linewidth]{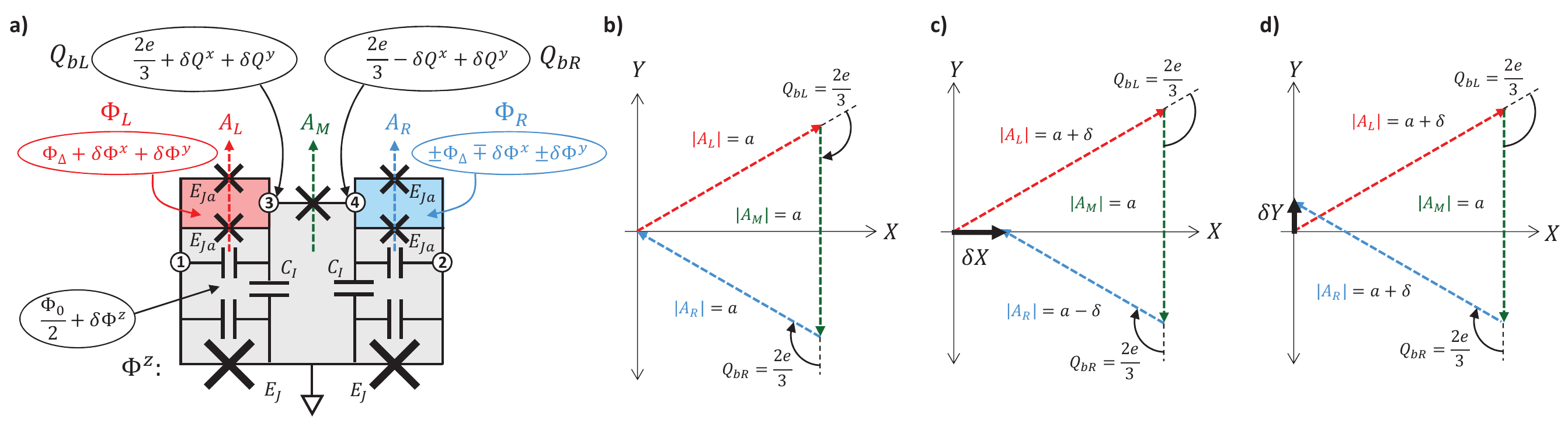}
    \caption[]
        {Emulating a Heisenberg quantum spin-1/2 with a JPSQ circuit. Panel(a) shows a circuit having two islands, and therefore three fluxon tunneling paths. The third tunneling path is realized with a single, fixed junction, while the other two are flux-tunable DC SQUIDs. The latter are biased with static flux offsets such that their fluxon tunneling amplitudes have flux sensitivities of equal magnitude. Panel (b) shows how by choosing the island charge offsets to be one third of a Cooper pair, the three tunneling amplitudes can be chosen to form an equilateral triangle in the complex plane, resulting in complete cancellation of tunneling. Panels (c) and (d) show how small changes to the two DC SQUID fluxes can then correspond to two orthogonal magnetic moments, via the common and differential modes of the two fluxes.}
        \label{fig:twoislJPSQ}
    \end{center}
\end{figure*}

In addition to the spurious couplings of noise inside the logical computational space, we must also consider processes in which the system is excited out of this space, requiring the environment to supply a photon at an energy equal to that of the energy barrier [c.f., $\Delta E_L$ in fig.~\ref{fig:BSqubitplots}(a)]. Although these processes are suppressed by the low temperature, they are also ``allowed" transitions for single photons coupling locally to the circuit, as opposed to the couplings inside the computational space just discussed, which only occur due to spurious non-idealities in the circuit realization of the Hamiltonian of fig.~\ref{fig:BSqubit}(a). Figures ~\ref{fig:BSqubitplots}(d) and (e) show the effective total rms dipole moments for these transitions from the two computational states to all of the states in the lowest excited manifold. As expected, these transitions are much stronger. We can readily estimate the resulting lifetimes of the logical states due to these transitions using the following flux and charge noise amplitudes at the frequency corresponding to the barrier height $\Delta E_L/h\sim2.1$ GHz, derived from the work of ref.~\onlinecite{MITLLflux}: $S_Q($2.1 GHz$)\sim(7.1 \times10^{-9}e)^{2}$/Hz, and $S_\Phi($2.1 GHz$)\sim(6.4\times10^{-11}\Phi_0)^{2}$/Hz. Assuming a thermal environment at $T=$20 mK so that the rates of absorption and emission processes at this frequency are related by the Boltzmann factor, we find in the region near zero logical field, a lifetime of $\sim$ 10 \textit{seconds} due to charge noise, and 1.5-6 \textit{milliseconds} due to flux noise. The latter, in fact, appears to be the coherence-limiting process for the parameters we have chosen, though it can likely be further improved by additional optimization of circuit parameters beyond that carried out here, with the specific goal of reducing the influence of these high-frequency magnetic transitions.

To employ such logical spins in a quantum annealing machine requires static, pairwise, logical Ising interactions between them. These would correspond to physical \textit{four-qubit} static interactions, which  have not yet been demonstrated experimentally. However, due to their relevance for applications such as adiabatic topological quantum computing \cite{adiatopo} and adiabatic quantum chemistry simulation \cite{adiaQChem,*bravyikitaev,*molecularenergies}, there are already several concrete proposals for realizing them \cite{jordanfarhi,*warburton,*GFCfourqubit,*schondorf}. A protected qubit like that shown in fig.~\ref{fig:BSqubit} could also be used in gate model applications. In the simplest case, the logical field control could be used to implement single qubit rotations (via Larmor precession around the emulated vector magnetic field), though two-qubit gates would still require four-physical-qubit interactions in some form. However, since only pulsed operations are required in a gate model context, they would not need to be static nor very strong. One possibility would be to use the techniques described in ref.~\onlinecite{AJKlongit}, which are based on state-dependent geometric phases accrued by the system when coupled to a resonator and driven, and which are readily scalable to multi-qubit entangling interactions. In this mode of operation, the logical field controls would no longer be necessary (reducing the magnetic dipole moment for physical noise to the $\sim$pA level, as mentioned above), and single and two-logical-qubit gates would be realized via selectively modulating the couplings between either two or four physical JPSQs, respectively, and a common resonator.

\section{Two-island JPSQ circuit for emulation of a 3D magnetic moment}\label{s:twoisl}

So far, we have discussed a JPSQ circuit that can be said to emulate a quantum spin-1/2, insofar as it has three physical operators that can be engineered to obey the canonical commutation relations in the computational subspace of the lowest two energy levels, while higher levels are kept relatively far away. However, only two of these operators are magnetic, and the remaining one is electric. In the case of fig.~\ref{fig:BSqubit}, only strong, fixed electric interactions are needed, which can be realized with the simple capacitive couplings between islands shown in fig.~\ref{fig:BSqubit}(b). However, in some of the applications mentioned in the introduction, one needs tunability of the couplings between all three components of the spins' dipole moments. Although there are proposed methods for engineering tunable electric couplings between superconducting circuits \cite{averin}, they tend to require the qubits to have sufficiently large electric dipole moments that charge noise is likely to become a major problem. Therefore, it would potentially be of great interest to realize a JPSQ circuit whose magnetic moment has three independent vector components that satisfy the spin-1/2 commutation relations, so that the well-established techniques for controllable magnetic coupling could be brought to bear to realize fully-controllable, anisotropic Heisenberg interactions.

Such a circuit is shown in fig.~\ref{fig:twoislJPSQ}. Here, we have added an additional fluxon tunneling path by including a third Josephson junction in the loop, and we bias the resulting two islands using separately-controllable voltages. As illustrated in fig.~\ref{fig:twoislJPSQ}, if the polarization charges on these two islands are both set to one third of a Cooper pair, and the three tunneling paths have the same amplitude, a completely destructive, three-path Aharonov-Casher interference again occurs, as in the two-path case discussed so far. Because of the $2\pi/3$ relative phase shifts between paths, we need only adjust two of the three amplitudes to control two (emulated) orthogonal transverse field directions, which can be accomplished using differential and common mode bias fluxes coupled to the two DC SQUID loops \footnote{As in the circuit of fig.~\ref{fig:JPSQfig1}, where a single island created both a magnetic and an electric transverse dipole moment, the two-island circuit of fig.~\ref{fig:twoislJPSQ}(a) has two orthogonal transverse electric dipole moments (in addition to the desired transverse magnetic dipole moments), corresponding to the common and differential modes of the two island voltages.}.

\section{Conclusion}\label{s:conclusion}

In this work, we have proposed for the first time a superconducting qubit circuit capable of emulating a spin-1/2 quantum object with a true, static, \textit{vector} dipole moment, whose components can be chosen, to a large extent, by design. We analyzed this circuit in detail, providing some qualitative intuition for its basic properties, and validated this with full numerical simulation of its quantum Hamiltonian. Broadly speaking, our results indicate that JPSQs with reasonable design parameters, when compared to existing flux and transmon qubits, can have a comparable or lower sensitivity to high frequency noise, but a somewhat higher sensitivity to low-frequency noise, with the latter almost entirely attributable to (and a necessary consequence of) their qualitatively more complex emulation capabilities. The most important open question, for the coherence of JPSQ, is whether its parity lifetime can be sufficiently increased using the bandgap engineering techniques that have been demonstrated so successfully in other experiments \cite{parityDiCarlo,*paritymarcus,*parityLeo}. If so, it would pave the way for high-coherence emulation of quantum spin-1/2 systems with nearly arbitrary pairwise vector interactions.

This capability would then enable the experimental exploration of a number of potentially significant ideas that until now have remained out of reach of available physical qubit hardware. These include, for example: the use of strong, non-stoquastic driver Hamiltonians in quantum annealing \cite{hormoziNS,*henconstrained,*nishimoriNS,*lidarNS,*crossonNS,*albashNS,*araiNS,*terhalNS} and quantum simulation \cite{simulation,*bosonsampling,*spinbosonsampling,Dwave3DTFIM,*DwaveKT}; Hamiltonians required for Hamiltonian and holonomic computing paradigms \cite{Lloydterhal,*cellularautomata,*hamiltonianQC1D,*holonomic,marvianlidar}; and engineered emulation of the full quantum Heisenberg model. If combined with one or more of the proposed schemes for realizing static \cite{jordanfarhi,*warburton,*GFCfourqubit,*schondorf} or pulsed \cite{AJKlongit} multiqubit interactions, an even wider range of possibilities would become accessible, including adiabatic topological quantum computation \cite{adiatopo}, adiabatic quantum chemistry \cite{adiaQChem,*bravyikitaev,*molecularenergies}, and quantum error suppression \cite{jiangrieffel,marvianlidar}.

\acknowledgments

This research is funded by the Office of the Director of National Intelligence (ODNI), Intelligence Advanced Research Projects Activity (IARPA), and by the Assistant Secretary of Defense for Research \& Engineering under Air Force Contract No. FA8721-05-C-0002. The views and conclusions contained herein are those of the authors and should not be interpreted as necessarily representing the official policies or endorsements, either expressed or implied, of ODNI, IARPA, or the U.S. Government.

\providecommand{\noopsort}[1]{}\providecommand{\singleletter}[1]{#1}%
%


\end{document}